\documentclass[preprint2]{proto}
\usepackage{times}
\usepackage{xcolor}
\usepackage{graphicx}
\usepackage{wasysym}
\usepackage{float}
\usepackage{dblfloatfix}

\graphicspath{{./}{Figures/}}

\begin{document}

\title{\textbf{\LARGE DYNAMICAL POPULATION OF COMET RESERVOIRS}}

\author {\textbf{\large N. A. Kaib}}
\affil{\small\em University of Oklahoma}

\author {\textbf{\large Kathryn Volk}}
\affil{\small\em University of Arizona}

\begin{abstract}

\begin{list}{ } {\rightmargin 1in}
\baselineskip = 11pt
\parindent=1pc
{\small 
The Oort cloud and the scattered disk are the two primary reservoirs for long-period and short-period comets, respectively. 
In this review, we assess the known observational constraints on these reservoirs' properties and their formation. 
In addition, we discuss how the early orbital evolution of the giant planets generated the modern scattered disk from the early, massive planetesimal disk and how $\sim$5\% of this material was captured into the Oort cloud. 
We review how the Sun's birth environment and dynamical history within the Milky Way alters the formation and modern structure of the Oort cloud. 
Finally, we assess how the coming decade's anticipated observing campaigns may provide new insights into the formation and properties of the Oort cloud and scattered disk. 
\\~\\~\\~}
\end{list}
\end{abstract}  

\section{\textbf{DISCERNING THE COMET RESERVOIRS}}
\label{sec:intro}

Comets make up one of the major populations of small bodies in the solar system, and their passages near Earth and the Sun have been documented by astronomers for millennia \citep{kronk99}. 
However, these bodies are known to have relatively short lifespans both dynamically, due to orbital instability, and physically, due to the volatile loss and surface dust deposition that occurs during each near-Sun perihelion passage \citep{weiss80, rick90, levdun94}. 
As a result, it has long been assumed that the short-lived modern population of comets are supplied from more populous distant reservoirs whose members have much longer stability timescales.

Until 1992, Pluto was the most distant known object in the solar system, and the nature of comet reservoirs had to be inferred solely from the properties of the more proximate but transient population of comets passing near Earth. Historically, the catalog of known comets has been broadly divided into two groups based on orbital period: short-period comets (SPCs) whose orbital periods are less than 200 years, and long-period comets (LPCs) whose orbital periods are greater than 200 years. It has  long been recognized that the orbital inclinations of comets with large periods are nearly isotropic, while short-period comets have an inclination distribution strongly concentrated toward prograde ecliptic orbits \citep[e.g.][]{everhart72, marsden2009}. 
This fundamental difference in the orbital distribution of SPCs and LPCs has fueled decades of dynamical studies attempting to understand how the supply of comets in the inner solar solar system is replenished and what the source reservoirs of this replenishment must look like (see, e.g., historical reviews by \citealt{Davies:2008,Fernandez:2020}).

\subsection{An Oort Comet Cloud}
Amongst the long-period comets, the distribution of the inverse semimajor axes ($1/a$) of the comets 
is a representation of their orbital energy distribution, and \citet{oort50} noted an overabundance of comets with $0<1/a < 0.0001$ au$^{-1}$, or $a>10^4$ au. \citet{oort50} posited that the comets belonging to this overabundance (now known as the Oort spike) are bodies that are making their first perihelion passage near the Sun, while long-period comets with smaller semimajor axes are on subsequent passages, their orbital energies having been drawn down via perturbations from Jupiter during previous perihelion passages. 
Jovian perturbations can also increase the orbital energies of new long-period comets during their initial perihelion passage, but these comets are so weakly bound that this process permanently ejects them from the solar system on hyperbolic orbits. 

To explain a continuous supply of new long-period comets, \citet{oort50} proposed that the solar system is surrounded by a vast cloud of $\sim$10$^{11}$ comet-sized bodies extending to $\sim$10$^5$ au. 
In this scenario, LPCs passing near Earth on extreme eccentricities comprise a tiny fraction of the bodies within this cloud, whose orbits have semimajor axes of at least thousands of au but whose eccentricities range all the way from 0 to 1. 
If this cloud were dynamically inert, its high-eccentricity (low-perihelion) portion should have been depleted long ago due to hyperbolic ejection events during perihelion passages. 
However, at such large distances from the Sun, perturbations external to the solar system become important. 
\citet{opik32} noted that impulses from passing stars will alter the perihelia (and therefore eccentricities) of distant heliocentric orbits. 
This effect allows new Oort cloud bodies to be brought into the inner solar system as previous long-period comets are lost through ejection by the giant planets, namely Jupiter. 
Moreover, these same perturbations will randomize the directions of orbital angular momenta, leading to an isotropic distribution of LPCs and a spherical structure to the cloud. 
Decades later it was realized that the tide of the Milky Way's disk (generated by the vertical distribution of matter within the disk) will also provide perihelion and inclination shifts in Oort cloud bodies that are, on average, larger than those due to passing stars \citep{mormul86, heitre86}. 
This perturbative force will also steadily replenish the supply of LPCs from lower eccentricity Oort cloud bodies. 

\subsection{The Edgeworth-Kuiper Belt and Scattered Disk}\label{ss:introSD}

While the seminal work of \citet{oort50} outlined most of the broad features of LPC dynamics and their distant reservoir, the origin of SPCs still remained a puzzle due to their lower-inclination orbits. 
\citet{everhart72} argued that the observed SPC population could be produced from multiple rounds of jovian perturbations that lower the semimajor axes of low-inclination LPCs. 
However, such a process appeared to be orders of magnitude too inefficient \citep{joss73, krespitt78} and would require an extremely massive Oort cloud \citep{fern80} to explain the abundance of SPCs. 

Meanwhile, it had long been speculated that the falling surface density of solid material with increasing heliocentric distance in the early outer solar system would produce a belt of icy planetesimals that were never incorporated into fully formed planets \citep{edge43, edge49, kuip51}. 
While \citet{kuip51} assumed that perturbations from Pluto (whose mass was then thought to be comparable to Earth's) had dispersed this belt into the Oort cloud, \citet{edge43} proposed that members of this belt occasionally visit the inner solar system as comets. 
In another study, \citet{everhart77} found that some Oort cloud bodies with perihelia near Neptune will eventually be scattered into Jupiter-crossing orbits and evolve into SPCs. 
\citet{fern80} realized that a belt of small bodies residing beyond Neptune would naturally supply near-Neptune perihelia and could generate a population of SPCs on its own.
However, it was assumed that a reservoir completely comprised of Neptune-encountering orbits would be quickly depleted over the age of the solar system due to its inherent orbital instability.

\citet{dqt88} used numerical simulations to show that the inclinations of Oort cloud bodies are approximately conserved as their semimajor axes are decreased via encounters with the giant planets, and they concluded that the Oort cloud cannot be the main source of SPCs. 
An additional set of simulations showed that a low-inclination reservoir of Neptune-crossing bodies generates a low-inclination population of SPCs similar to the observed population, and these results were refined and confirmed in \citet{qtd90}. 
Thus, it seemed that the population of SPCs required the existence of a second reservoir of bodies in a belt beyond Neptune. 

The mechanism that injected members of this belt into the comet population was still unclear, however. 
Planetesimal-planetesimal scattering suggested by \citet{fern80} required such massive belt members that they would not have evaded detection. 
\citet{Torbett:1989} showed instead that chaos resulting from the gravitational effects of the four giant planets could raise the eccentricities of initially more circular orbits to slowly supply Neptune-encountering bodies, and therefore comets, from this region.
Direct integrations showed that a similarly slow destabilization process takes place near the boundaries of mean motion resonances with Neptune \citep{dunlev95}. 
However, additional work in \citet{levdun97} found that most regions of orbital space of the conventional Kuiper belt (low-eccentricity orbits from $\sim$40-50 au) either destabilize quickly ($\lesssim$10$^8$ years) or are stable over timescales longer than the age of the solar system. 
However, \citet{dunlev97} noted that a small subset ($\sim$5\%) of Kuiper belt objects that have a close encounter with Neptune will attain $\sim$Gyr dynamical lifetimes. 
They achieve these long dynamical lifetimes by occupying eccentric orbits whose perihelia  have been temporarily lifted away from Neptune, typically within mean motion resonances. 
With a dynamical lifetime comparable to the solar system's age, these ``scattered disk'' bodies are a more efficient producer of SPCs compared to the classical Kuiper belt. 
Thus, the subpopulation of trans-Neptunian objects (TNOs) known as the scattered disk is widely thought to be the main source of SPCs. 

\subsubsection{Centaurs as a Comet-Reservoir Link}

Before TNOs make it all the way into the inner solar system, they must pass through the giant planet region as Centaurs, the intermediary small body population connecting short-period comets and the scattered disk (e.g., \citealt{dunlev97}; reviewed recently by \citealt{Peixinho:2020}).
The exact orbital definition of Centaurs differs in the literature \citep[e.g.][]{Gladman:2008,Jewitt:2009}, but the general criteria is that they are small bodies on giant-planet crossing orbits that are dynamically unstable on timescales much shorter than the age of the solar system. 

The first discovered object widely recognized as belonging to the new class of objects later dubbed the Centaurs was (2060) Chiron \citep{Kowal:1977}.
It was quickly realized that Chiron's orbit is dynamically unstable \citep[e.g.][]{Oikawa:1979,Scholl:1979}.
More general studies of orbital stability in the giant planet region reveal that Centaurs have typical dynamical lifetimes on the order of 1-10 Myr \citep[e.g.][]{Gladman:1990,Holman:1993,tismal03,DiSisto:2020}.
Centaurs thus are a transition population for comets en-route from the transneptunian region \citep[e.g.][]{Stern:1996} rather than a reservoir.
However, not all Centaurs become comets.
Centaurs spend most of their time in the outer, ice giant region, with only about half making it to the Jupiter-Saturn region and about a third evolving onto short-period comet orbits \citep[e.g.,][]{tismal03,Sarid:2019}.
The chapter by Fraser et al. in this volume discusses the evolution of Centaurs into comets in more detail.

 

\subsection{Bodies with Ambiguous Sources}

There are a few groups of comets/comet predecessors that do not fall neatly along the long- or short-period divide and are not easily produced by models of Oort Cloud or scattered disk delivery.
In addition to orbital period, comets are often divided into subgroups based on their Tisserand parameter with respect to Jupiter, $T_J$:
\begin{equation}
    T_J = \frac{a_J}{a} + 2\sqrt{\frac{a}{a_J}(1-e^2)}\cos{i},
\end{equation}
where $a_J$ and $a$ are the semimajor axes of Jupiter and the comet, respectively, $e$ is the eccentricity of the comet's orbit, and $i$ is its orbital inclination.
This parameter is related to a comet's Jacobi constant, which is a strictly conserved quantity in the restricted three-body problem (Sun + circular Jupiter + comet) and is approximately conserved on 1000-year timescales for SPCs in the real solar system (\citealt{levdun94}; see also, e.g., \citealt{SSD1999} for a full discussion of the Jacobi and Tisserand parameters). 
It has long been noted that the vast majority of short-period comets have $T_J$ slightly below 3, indicating a tight dynamical coupling with the gas giant \citep{kresak72}, and \citet{levdun97} demonstrated how dynamical transfer from the scattered disk generates this tight distribution of $T_J$. 
Based on these results, \citet{lev96} argued that the Tisserand parameter should be the main criterion upon which to base comet classification. 
In particular, \citet{levdun97} argue that comets with 2 $<T_J<$ 3 should be considered Jupiter-family comets and are consistent with a Kuiper belt origin. 

\begin{figure}[htb]
    \centering
    \includegraphics[scale=1.35]{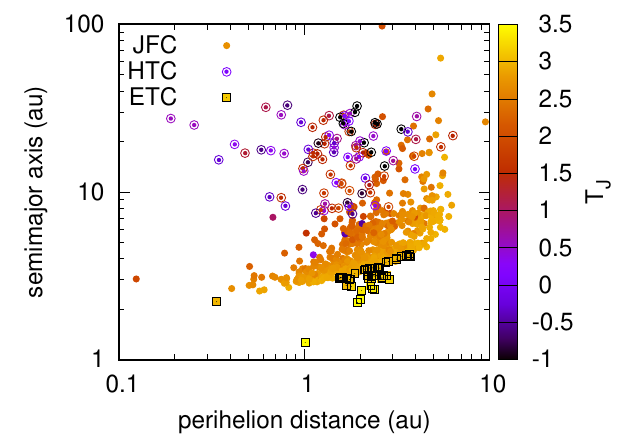}
    \caption{Semimajor axis vs perihelion distance for all Jupiter-family (JFC; circles), Halley-type (HTC; triangles), and Encke-type (ETC; squares) comets listed by JPL (queried from https://ssd.jpl.nasa.gov/sbdb\_query.cgi on Aug. 22, 2021). The comets are color-coded by Tisserand parameter with respect to Jupiter ($T_J$) to highlight the separations between the different dynamical classes of short-period comets.}
    \label{fig:comets}
\end{figure}

\subsubsection{Halley-type Comets}\label{sss:htc}
Figure~\ref{fig:comets} shows the current known population of SPCs, color coded by $T_J$; it is clear that not all comets are confined in the tight cluster of $T_J=2-3$.
In particular, there exists a subgroup of SPCs with $T_J < 2$, which \citet{carusietal87} designated Halley-type comets (or HTCs). 
The driver for their lower $T_J$ values are their generally significantly larger orbital inclinations (up to and including retrograde orbits) as well as moderately larger semimajor axes.
The ultimate source of these comets has been debated, with \citet{lev01} suggesting that HTCs are consistent with an origin from a flattened interior of the Oort cloud, \citet{lev06} arguing for a scattered disk origin whose most distant members are pushed inward via the Galactic tide (which also isoptropizes ecliptic inclinations), and \citet{Nesvorny:2017} arguing that HTCs are consistent with returning LPCs from the Oort cloud whose semimajor axes have been drawn down through repeated planetary encounters.
\cite{Fernandez:2018} find that HTC orbits can be produced from integrations of inactive Centaurs with perihelia in the Jupiter-Saturn region, suggesting that the lack of observed cometary activity could be due to a long residence time at these heliocentric distances that depleted surface volatiles. 
While \cite{Fernandez:2018} assume a scattered disk origin for all the Centaurs, perhaps this is also consistent with the \cite{Nesvorny:2017} returning LPC scenario.
Additional modeling and observations are likely necessary to test these different proposed HTC origins.

\subsubsection{Encke-type Comets}

On the opposite side of the $T_J$ divide from the JFCs are the Encke-type comets (ETCs); these comets typically have $T_J$ just above 3 with low semimajor axes that slightly de-couple the comets from Jupiter compared to JFCs (see Figure~\ref{fig:comets}). 
\cite{Levison:2006} investigated the origins of comet Encke and found that such orbits can be part of the standard model for JFC evolution, but that it is a rare outcome that typically required $1.5\times10^5$ perihelion passages in the inner solar system to achieve. 
This long timespan is problematic for producing active comets like the observed ETCs because comets are expected to fade (either via disruption or loss of volatiles) on timescales of order a few tens to a few hundred perihelion passages \citep[e.g.][]{levdun97,brassschwamb15}.
\cite{Levison:2006} suggest that Encke could perhaps have built up an inactive surface layer during this time which was then shed after a dramatic drop in $q$, but it remains unclear whether this kind of evolution can plausibly explain the entire ETC population.
It is possible that the production of ETCs depends on more than just purely gravitational dynamics. 
Most dynamical models of how short period comets are supplied from their source regions neglect nongravitational forces such as outgassing, but these might be important for the ETCs. \cite{Fernandez:2002} found a higher rate of ETC production when such forces were included in the model, though they found they needed to be sustained over a relatively long timeframe.
Future models of ETC production with more realistic nongravitational forces would likely help shed light on this question.

\subsubsection{Highly Inclined Centaurs}

While the majority of Centaurs have low-to-moderate inclinations consistent with production from the scattered disk en-route to the JFC population, there exist a number of very high-inclination and retrograde Centaurs and TNOs with perihelia in the giant planet region \citep[e.g.][]{Gladman:2009,Chen:2016}. 
Much like the high-inclination HTCs, the origin of these highly inclined objects is unclear.
It has been proposed that such bodies could originate from the Oort cloud \citep{bras12, kaib19}. 
In this scenario, galactic perturbations drive the perihelia of Oort cloud bodies (whose inclination distribution is nearly isotropic) into the giant planet region at which point energy kicks from planetary perturbations can draw these bodies' semimajor axes to smaller values \citep{emel05, kaib09}. 
An alternative explanation for high-inclination centaurs is that they are signature of a distant undetected planet that perturbs the inclinations of scattered bodies \citep{gom15, batbrown16b}. 
However, such a mechanism may generate more high-inclination centaurs than are actually observed \citep{kaib19} and has other, observable consequences for the distribution of TNOs \citep{Shankman:2017}.
Another recently proposed origin for this population is outward scattering of objects from the asteroid populations in the inner solar system, which can produce very high-inclination and retrograde orbits \citep{Greenstreet:2020}. 
This last hypothesis would have interesting (and testable) compositional implications for the highest-inclination outer solar system objects.

\section{Observational Constraints On Reservoirs}

\subsection{Short-Period Comet Population}

\begin{figure}[htb]
    \centering
         \includegraphics[width=0.45\textwidth]{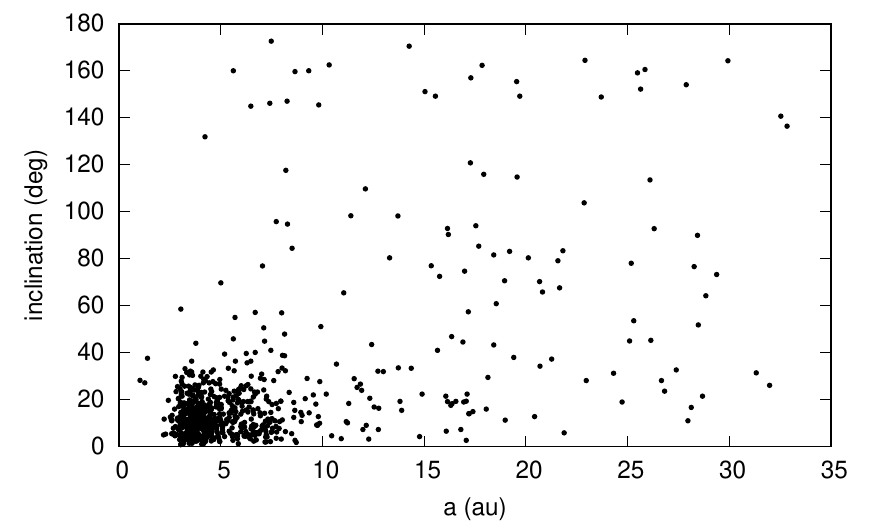} 
    \caption{Inclination vs semimajor axis distribution of observed cometary objects with orbital periods shorter than 200 years (the standard definition of SPCs; queried from JPL). The concentration of comets with low-$a$ and low-$i$ orbits highlight the JFC population.}
    \label{fig:obs-spcs}
\end{figure}

Figure~\ref{fig:obs-spcs} shows the current inventory of solar system objects with orbital periods less than 200 years classified as comets by JPL. 
There are obviously observational biases in this sample, but it highlights a few important features of the SPC population. 
First, note the large spread of inclinations for objects with $a \gtrsim 5$~au; many of the high-inclination objects in this group are the HTCs discussed in Section~\ref{sss:htc}, whose dominant source population remains an active area of research.
Second, note that the vast majority of comets with semimajor axes near or inside Jupiter's orbit (the JFCs) have inclinations smaller than $\sim30^\circ$.
This concentration of inclinations relatively close to the ecliptic plane was the first notable evidence linking JFCs to the outer solar system's TNO populations (as discussed in Section~\ref{ss:introSD}), which have a similar inclination distribution.

If we knew the exact number of SPCs in the inner solar system at the moment, we could use our dynamical models of the delivery of these comets to constrain the total number of comet-sized TNOs in the scattering population (the dominant supplier of these comets). 
However, this extrapolation is complicated by the fact that our observed sample of SPCs is biased toward active comets on easy-to-observe orbits. 
The bias toward active comets is particularly important because we have strong evidence that SPCs must `fade' (i.e., either become inactive and faint or be physically destroyed) on timescales much shorter than their dynamical lifetimes. 
The inclination distribution of the JFCs is the dominant evidence for this. 
Over their entire $\sim0.5$~Myr dynamical lifetimes in the inner solar system \citep{levdun94}, interactions with the planets tend to raise the orbital inclinations of JFCs to values larger than those seen in the observed population. 
To get the modeled JFC population to match the observed one in inclinations, \cite{Brasser:2015jfc} estimate that comets could fade after as few as $\sim50$ perihelion passages; \cite{Nesvorny:2017} model fading as a size-dependent process and estimate that small comets remain visible for a few hundred perihelion passages (both models consider perihelion passages with $q<2.5$~au). 
Our models of comet fading are thus still quite uncertain, making it difficult to extrapolate the current census of active JFCs into a total population of active and inactive JFCs.
Additionally, the very fact that comets are active makes measuring their sizes more difficult because the gas and dust coma obscures the comet nucleus. 
Thus measurements of the size distribution of comets, necessary for connecting population estimates to their source region, can vary \citep[e.g.][]{Lowry:2008,Snodgrass:2011,Fernandez:2013}; this, combined with different assumptions about the ratio of active to dormant comets, means estimates of the current number of JFCs can vary by factors of several (see, e.g., discussion in \citealt{Nesvorny:2017}).
These variable comet population estimates combined with variations in orbital models for their scattering population reservoir means that attempts to extrapolate from the number of observed JFCs to the number of comet-sized objects in the TNO region can differ enormously. 
Published estimates of the current population of the scattered disk based on supplying the current number of JFCs include: $6\times10^8$ ('comet-sized'; \citealt{dunlev97}), $\sim10^{10}$ ($D>1.4$~km; \citealt{Emel'yanenko:2004}), $0.5-1.1\times10^9$ ($D>1$~km; \citealt{Volk:2008}), $\sim6\times10^9$ ($D>2.3$~km; \citealt{Brasser:2015jfc}), $4.4\times10^8$ ($D>2$~km; \citealt{Nesvorny:2017}).
While these estimates are not all for identical size ranges, it is clear that they vary by at least an order of magnitude.

\subsection{Direct Observation of the Scattered Disk}\label{ss:obsSD}

\begin{figure}[htb]
    \centering
    \begin{tabular}{c}
        \includegraphics[width=0.47\textwidth]{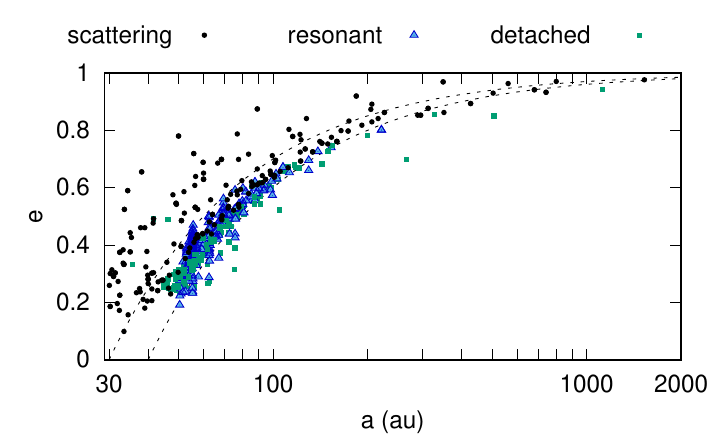} \\
        \includegraphics[width=0.47\textwidth]{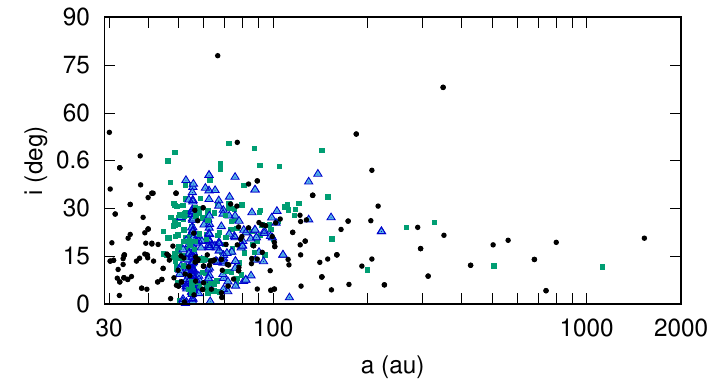} \\
    \end{tabular}
    \caption{Observed distribution of scattering, detached, and large-$a$ ($a>50$~au) resonant TNOs in eccentricity (top panel) and inclination (bottom panel) vs semimajor axis. The dashed lines in the top panel show constant perihelion distances of 30 and 40 au for reference. We include only objects with known dynamical classifications (we use the set of classified objects listed in \citealt{Smullen:2020}) to highlight the overlaps between these three dynamical classes (see \citealt{Gladman:2008} for detailed definitions of these classes).
    }
    \label{fig:obs-tnos}
\end{figure}

Since the detection of the first scattering TNO \citep{Luu:1997}, hundreds of additional members of this population and the closely related high-$a$ resonant and detached populations have been detected. 
Our picture of the high-$a$ TNO population has evolved as observations built up.
It became apparent fairly early that not all of the high-$a$ TNOs had perihelion distances that brought them into direct gravitational range of Neptune; observations of objects with high perihelion distances ($q\gtrsim40$~au) led to discussion of an `extended' scattered disk \citep[e.g.][]{glad02} or a dynamically `detached' population (the current more commonly used terminology; e.g., \citealt{Gladman:2008}).
The important role of Neptune's resonances in the scattering population was noted in early dynamical models \citep{levdun97}, and their potential role in creating the detached population was also noted \citep[e.g.,][]{gom05,Gallardo:2006}. 
As the orbits of distant TNOs have been measured more precisely, it is also apparent that there are large, stable populations of objects in many of Neptune's distant resonances \citep[e.g.][]{Gladman:2012,Pike:2015,Volk:2018} in addition to a large number of metastable objects temporarily stuck in resonances on a variety of timescales \citep[e.g.][]{Lykawka:2007,Bannister:2016r,Holman:2018}.
Figure~\ref{fig:obs-tnos} shows the observed orbital distribution of TNOs in the scattering, $a>50$~au resonant, and detached populations.

It was apparent in the early TNO datasets that the higher-eccentricity TNOs (including the scattering objects) have a much wider range of orbital inclinations than the low-eccentricity `classical' Kuiper belt objects, giving the total TNO population a bimodal inclination distribution \citep{Brown:2001}. 
This gave rise to a division of the TNOs into dynamically `cold' and `hot' components, with the cold population being found only in the classical belt.
The scattering, resonant, and detached populations all belong to the hot population.
This division has been strengthened by subsequent surveys of the physical properties of TNOs.
The photometric colors of hot and cold TNOs are distinct \citep[e.g.][]{Tegler:2003,Pike:2017,Schwamb:2019}, implying that they have different surface compositions, perhaps reflecting different formation regions.
The cold population also contains significantly more binary objects than the hot population; the few observed binary objects in the hot population tend to have a massive primary with small satellites rather than the prevalent equal-sized binaries in the cold population (see a recent review by \citealt{Noll:2020}).
There was also early evidence that the hot population's size distribution differed from the cold population.
\cite{Bernstein:2004} found evidence for this difference, finding significantly fewer hot population objects at faint magnitudes than expected. 
Subsequent observational surveys have found additional evidence for differences between the hot and cold populations
and for a `turnover' in the magnitude/size distribution from a steep slope for the bright/large objects to a shallower one at fainter/smaller objects 
\citep[e.g.,][]{Fraser:2009,Fuentes:2009,Fraser:2010,Shankman:2013,Fraser:2014,Kavelaars:2021}; 
however the exact nature and location of this change has been difficult to constrain due to the strong observational biases and faintness of the TNOs.
Recent results from the Outer Solar System Origins Survey (OSSOS) have shown that the scattering population of TNOs can be well fit with either a break or a `divot' (an actual decrease in the differential $H$ magnitude distribution rather than just a change in slope) near $H_r\approx8-8.5$ \citep[e.g.][]{Shankman:2013,Shankman:2017,Lawler:2018}.

All of these observations have led to the current view that the hot populations, which are the progenitors of the short period comets, formed closer to the sun than the current cold classical region. 
They formed in the current giant planet region and were dispersed onto their current orbits during the epoch of giant planet migration (discussed in Section~\ref{sec:building}).
At present, observational estimates of the number of comet-sized objects in the scattering population are still fairly uncertain. 
In reflected sunlight, km-sized bodies in the scattering population are far too faint for direct detection even at perihelion. 
Comet-sized population estimates thus must be extrapolated based on the size distribution of brighter bodies.
But even the population estimates for these brighter bodies have large uncertainties because the observational biases in the scattering population are so strong (see, e.g., discussion of observational biases in \citealt{Gladman:2021}).
Briefly, TNOs with large semimajor axes and large eccentricities spend most of their time at large heliocentric distances; they are typically only bright enough to be detected near their perihelia and thus over only a small portion of their orbits.
Estimating the intrinsic number of scattering objects from the observed sample of objects requires either weighting each observed object by the inverse of its probability of detection based on its size and orbit (a traditional `debiasing' approach) or taking a forward-modeling approach and applying observational biases to a model of the scattering population and iterating that model until it gives you a match to the observed sample. 
The accuracy of both approaches rely to a large extent on how well the biases of a given observational survey are understood and measured as well as how well the observed objects sample the orbital distribution of the scattering population (this is discussed further in Section~\ref{ss:rubin}).
\cite{Adams:2014} provide estimates of the scattered disk population based on debiasing the objects detected by the Deep Ecliptic Survey \citep{Elliot:2005}; they estimate that there are $\sim10^4$ scattered objects with absolute magnitudes $H>7.5$.
Taking a forward modeling approach, \cite{Lawler:2018} provide population estimates for the scattering population from OSSOS.
They use the \cite{kaib11} model of the intrinsic orbital distribution of the scattering population (with a slightly modified inclination distribution to better match the observed distribution, see \citealt{Shankman:2016}) with a variety of functional forms for the absolute magnitude distribution to find acceptable matches to the observed sample.
From this, \cite{Lawler:2018} estimate that there are $(0.7-1)\times10^5$ scattering objects with $H_r>8.66$ (approximately $D>100$~km; range covers all of their $H$ magnitude models); extrapolating down to $H_r<12$ ($D\gtrsim20$~km) yields population estimates of $\sim(2-3)\times10^6$ objects.
These are our current best estimates of the scattering population based on direct observations; while these estimates will likely improve with additional observations of the scattering population (see Section~\ref{ss:rubin}), the inherent brightness limitations of TNO surveys will make it difficult to push the direct constraints on the in-situ scattering population down to comet-sized bodies.

Two avenues of current research show promise for connecting the observational constraints on the size distribution of larger scattering objects to the population of comet-sized ones: occultation surveys (discussed in Section~\ref{ss:occultation}) and analysis of cratering records in the outer solar system.
In the latter case, data from the New Horizons mission provided a wealth of new data about the distribution of impact craters on TNOs Pluto and Arrokoth (see, e.g., \citealt{Singer:2019,Spencer:2020}).
Impact crater distributions do not easily translate to \textit{direct} constraints on the TNO size distribution. 
In addition to some uncertainties in crater scaling laws, converting crater sizes to impactor sizes requires knowing the impact speed distribution onto the surface in question, and this distribution depends on the assumed model for the orbital distributions and relative populations of the various dynamical classes of TNOs (see discussion in \citealt{morb21}); despite these uncertainties, the crater distributions offer our strongest current constraints on the distribution of the smallest TNOs.
The scattering population does not dominate the impact flux at either Pluto or Arrokoth \citep{Greenstreet:2015,Greenstreet:2019}, so the crater size distribution does not directly constrain its total population at small sizes.
However the hot classical population of TNOs (which is expected to have similar origins as the scattering population; see Section~\ref{sec:buildingSD}) does contribute to the cratering on both bodies, so the cratering record can offer insights into the size distribution of the scattering population even if not the absolute numbers; hopefully future detailed modeling can connect the implied combined TNO impactor distributions to population estimates for the comet-sized portion of the scattering population.

\subsection{Direct Observation of Oort Cloud}

With semimajor axes of thousands of au and perihelia decoupled from the solar system's planetary region, direct detection of Oort cloud objects is exceptionally difficult. To date, the TNO 2015 TG$_{387}$ \citep{Sheppard:2019} is the most likely candidate for a directly detected Oort cloud object (outside of the LPCs and other lower perihelia bodies whose orbits have been substantially modified by recent planetary interactions). With a perihelion of 65 au, it is clearly very decoupled from scattering events with the giant planets. Its semimajor axis is 1170 au, meaning that it is sensitive to the stellar and galactic tidal perturbations that influence the rest of the Oort cloud \citep{Sheppard:2019}. 

With a sample of just one detected object, it is extremely difficult to make confident inferences about the nature of the Oort cloud. Given the steep dependence of discovery probability on semimajor axis, we do not have strong constraints on the semimajor axis distribution of the Oort cloud. In addition, while 2015 TG$_{387}$ is estimated to have a diameter of 300 km, one cannot construct an accurate size distribution with a single object. Additional direct detections will be required to develop confident constraints on the Oort cloud's properties. 

\subsection{Long-Period Comet Flux}

Because of the extreme distance to the Oort cloud, 
the vast majority of observational constraints come from observations of LPCs. In particular, the flux of dynamically new ($a>10^4$ au) comets near Earth is one of the main observational parameters to constrain the cloud. The reason that comets with semimajor axes above $10^4$ au are considered dynamically new is that a single passage through the gas giant region will typically deliver an energy kick strong enough to either eject such a comet onto a hyperbolic orbit or lower it to a much smaller semimajor axis. 

For the last several decades, efforts have been made to measure the annual flux of dynamically new comets with the idea that such measurements will constrain the mass and population size of the Oort cloud. (A more massive or more populous cloud will generate larger numbers of LPCs.) While quantifying the flux, most works have sought to measure the number of new comets whose absolute magnitude, $H_T$, is brighter than some value. Here, $H_T$ is given by 

\begin{equation}
    H_T = m - 2.5n\log{r} - 5\log{\Delta}
\end{equation}
where $m$ is the total apparent magnitude, $r$ is the distance to the Sun, $\Delta$ is the distance to Earth, and $n$ is the comet's photometric index, or the power-law index defining the relationship between the comet's brightness and heliocentric distance. $H_T$ is the absolute magnitude of a comet (nucleus and coma) when it is 1 au from the Earth and Sun. For a non-active body $n=2$, but for a comet whose outgassing increases during a solar approach $n$ is often higher. (If one assumes $n=4$, as is often done, the absolute magnitude is noted as $H_{10}$.)

An oft-quoted metric is the flux of dynamically new comets per year with perihelia below 4 au (interior to Jupiter's orbit and potentially observable from Earth) and whose $H_T$ is brighter than 11. \citet{everhart67} found that $\sim$20 new LPCs pass within 4 au of the Sun per year, but this result relied on the extrapolation of an uncertain comet magnitude distribution derived from historical comet searches with selection biases that were extremely difficult to quantify. Meanwhile, utilizing a greater number of detections from modern surveys \citet{francis05} found a flux of only 2.9 per year, and \citet{fouch17b} found a flux of 3.6 per year. 

\subsubsection{Population of Oort cloud}

To estimate the total population of the Oort cloud, many past works have examined the new LPC production rate within simulations of the Oort cloud and scaled this value to the observed LPC flux. \citet{kaibquinn09} found that, on average, 1 in every $\sim$10$^{11}$ Oort cloud bodies passes through perihelion within 4 au of the Sun per year. In contrast, \citet{fouch14} found a lower rate of 1 in $5\times10^{11}$, but this may have been due to the higher level of central concentration in their Oort Cloud semimajor axis distribution, which placed more bodies onto small semimajor axes with lower comet production rates \citep{kaibquinn09}. Taking a mean of the two results and assuming a real flux of 3 new comets per year with $q<4$ au and $H_T<11$, yields a total Oort cloud population of 7--8 $\times10^{11}$ cometary bodies with $H_T<11$. 

\subsubsection{Mass of Oort cloud}

$H_T$ is of course not an actual direct measure of a comet's nucleus, because a comet's brightness is dominated by the coma when it is at 1 au. To discuss the flux of comets of an actual size or mass, $H_T$ has often been converted to a nuclear magnitude, which is then converted to a size/mass, but such a conversion is fraught with uncertainty \citep[e.g.][]{bail88, weiss96}. Using mass-$H_T$ relationships derived from observations of Halley's Comet in combination with the long-period comet $H_T$ distribution found in \citet{everhart67}, \citet{weiss96} determined the average long-period comet mass to be 0.8--$4\times10^{16}$ g. This would imply an Oort cloud mass of 1--5 M$_{\oplus}$. 

Alternatively, \citet{boe19} utilized Pan-STARRS1 long-period comet detections and applied a universal comet sublimation model to convert each cometary magnitude to nucleus radius. With this approach, they found that the LPC size distribution is a broken power-law with a steep bright-end index ($r^{-4.6}$) transitioning to a shallow faint-end index ($r^{-1.5}$) near an inferred comet radius of 2 km. This diminishes the relative contribution of sub-km bodies to the mass of the Oort cloud, and \citet{boe19} estimated a total Oort cloud mass of 0.5--2 M$_{\oplus}$ in bodies larger than 1 km in radius. 

\citet{sosfern11} took yet a different approach and searched for a relationship between the observed water sublimation and non-gravitational forces altering comets' orbits to find the masses of LPCs. Using this relationship, they then derive one for the total visual magnitude and mass of a comet:
\begin{equation}\label{eqn3}
    \log{R {\rm(km)}} = 0.9 - 0.13H_T.
\end{equation}
While this method was a unique and independent attempt to measure LPC nuclei, it relies upon several uncertain parameters, including an assumed bulk density of comets and a conversion of visual magnitude into water sublimation rate. The above mass-visual magnitude relationship was derived over a relatively narrow range of absolute magnitudes of 5--9, corresponding to radii of 0.5--1.8 km according to Eq.~\ref{eqn3}. Applying this conversion relation to a broader number of long-period comet discoveries, \citet{fernsos12} inferred a comet size distribution comprised of three different power laws. However, extending their bright-end power law to infinity implies an infinite mass to the Oort cloud, so it must become steeper at some larger comet nucleus size beyond the range they study ($r\sim$20 km).

\subsubsection{Population Ratio of the Oort Cloud to the Scattered Disk} \label{sec: ratio}

Historically, JFCs and LPCs have been used to estimate the population of comet-sized bodies within the scattered disk and Oort cloud, respectively. Taken at face value, these population estimates imply that there are nearly 3 orders of magnitude more such bodies in the Oort cloud than the scattered disk \citep{lev10}. However, these estimates still largely rely on uncertain conversions of $H_T$ to comet nuclear magnitudes. \citet{brasmorb13} attempted to limit the uncertainty by only considering LPCs and JFCs with inferred radii above 2.3 km, which largely avoided the previously discussed breaks in the LPC size distribution. With this approach, they found a substantially smaller Oort cloud-to-scattered disk population ratio of 44$^{+54}_{-34}$. In line with these results, \citet{Nesvorny:2017} and \citet{vok19} suggest that the sizes of LPC nuclei may be overestimated and are typically just 300-400 m in diameter, rather than the km sizes inferred in most works \citep[e.g.][]{francis05, fernsos12, boe19}. This would also lower the population of the Oort cloud and its ratio to the scattered disk \citep{Nesvorny:2017}. 

\section{\textbf{BUILDING THE COMET RESERVOIRS}}
\label{sec:building}

\subsection{A Scattered Oort Cloud}

\subsubsection{A Rough Framework}

Although the population of LPCs is very consistent with the existence of a distant spherical Oort cloud, they do not tell us how this cloud was formed. \citet{oort50} realized, however, that the same stellar perturbations that produce LPCs from the Oort cloud can also explain the cloud's formation. If solid planetesimals are leftover after the formation of the planets, it is inevitable that many will eventually undergo close encounters with the giant planets, exchanging energy with the planets and driving a diffusion in the planetesimals' semimajor axes. As the planetesimals' semimajor axes diffuse, their perihelia effectively remain locked in the giant planet region, allowing energy exchanges to occur during each perihelion passage. Left unchecked, this process would eventually result in all of the planetesimals either being ejected onto hyperbolic orbits or colliding with the Sun or a planet. 

However, if planetesimals reach semimajor axes of thousands of au, the perturbations from passing stars begin to become significant \citep[e.g.][]{rick76}. If stellar perturbations can torque the planetesimals' perihelia beyond the planetary region this would save these planetesimals from planetary ejections onto hyperbolic orbits. These same perturbations would act to isotropize the orbital planes of planetesimals and could give rise to a distant spherical cloud of planetesimals similar to that inferred from LPC observations. 

Thus, \citet{oort50} hypothesized a basic framework for the formation of the Oort cloud, but at the time, the efficiency of such a process was unknown. While \citet{oort50} assumed that the planetesimals occupying the Oort cloud originated between Mars and Jupiter, \citet{fern78} noted that the probability that planetesimal-planet encounters result in a near-parabolic orbit rather than a hyperbolic one increases with planet-Sun distance. As a result, material scattering off of Uranus and Neptune should be more likely to reach the Oort cloud than planetesimals closer to the Sun. This process was numerically verified in \citet{fern80b}, where particles were initiated on near parabolic orbits with perihelia between 20--30 au. They were numerically integrated as they scattered off of the giant planets while also being torqued by passing stars for 10$^9$ years. In this time span, over 50\% of bodies were ejected while most survivors resided in a nearly isotropized cloud between 20 and 30 thousand au. Extrapolating from these results, \citet{fern80b} concluded that Uranus and Neptune implanted $\sim$10\% of the material they scattered into the Oort cloud. 

Another early attempt to model the construction of the Oort cloud through these mechanisms was \citet{shoewolfe84}. \citet{shoewolfe84} assumed the planetesimals were leftover from the formation of Uranus and Neptune and employed a distribution of initially circular orbits between 15 and 35 au for their planetesimals. The evolution of these planetesimals' orbits due to planetary encounters was modeled with a statistical approach \citep{opik76}, and the influence of passing stars was implemented as in \citet{oort50}. Largely in agreement with \citet{fern80b}, \citet{shoewolfe84} found that $\sim$10\% of planetesimals will avoid ejection or collision with the Sun. Of those survivors, the vast majority ($\sim$90\%) will be retained on orbits with perihelia beyond the planets' orbits and with semimajor axes between 500 and 20,000 au. A smaller fraction ($\sim$10\%) were predicted to orbit with semimajor axes beyond 20,000 au, comprising the Oort cloud that \citet{oort50} predicted to explain LPC production. Thus, the survivors in the \citet{shoewolfe84} work tended to orbit the Sun at closer distances than the \citet{fern80b} work. 

\subsubsection{A Numerical Model}

\begin{figure}[htb]
    \centering
    \includegraphics[scale=0.55]{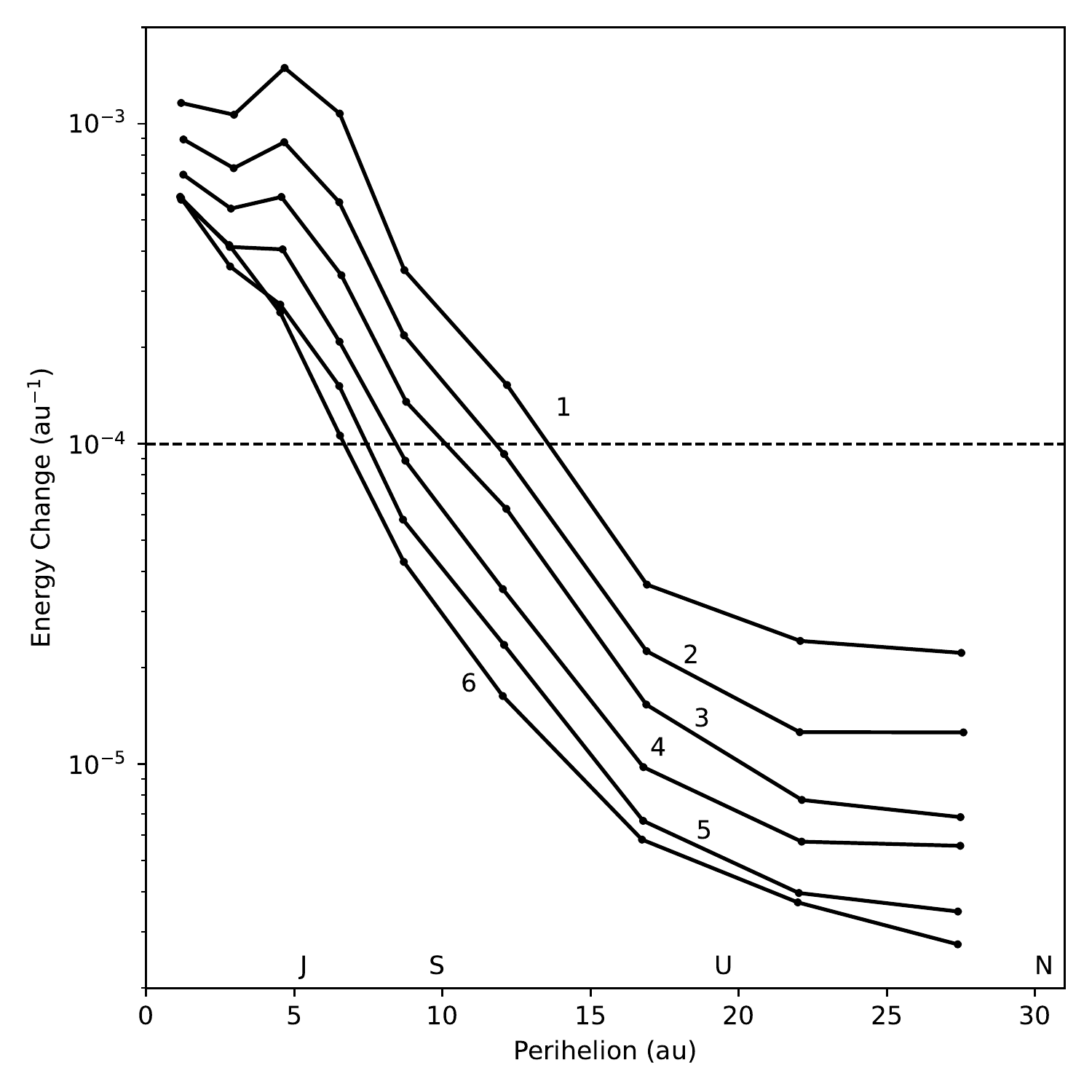}
    \caption{Typical energy change of a parabolic particle during a single perihelion passage as a function of perihelion. Solid lines 1--6 mark 30$^\circ$ inclination bins from 0--30$^{\circ}$ to 150--180$^\circ$. The dashed line coincides with the approximate width of the Oort cloud energy window for particles scattering off Jupiter and Saturn. Adapted from \citet{fern81}.}
    \label{fig:dEvsq}
\end{figure}

This process of Oort cloud formation was then revisited with still more direct numerical techniques in \citet{dqt87}. In this work, planetary perturbations on planetesimals were still applied with a Monte Carlo approach once per planetesimal orbit. However, the Monte Carlo sample was built from a distribution of energy kicks compiled from direct integrations of planetesimals through the giant planet region of the solar system. (For an example of such a distribution, see Figure \ref{fig:dEvsq}). As planetesimals reached large ($>$2000 au) semimajor axes, their orbits were directly integrated. During these direct integrations, planetary energy kicks were still applied at each perihelion passage, stellar perturbations consistent with the solar neighborhood kinematics were applied via the impulse approximation, and a vertical (with respect to the Galactic plane) tidal force due to the distribution of matter within the Milky Way's disk was included \citep{heitre86}. 

The approach of \citet{dqt87} revealed the key timescales involved that govern the formation of the Oort cloud. The semimajor axes of planetesimals diffuse due to planetary energy kicks received near perihelion. The timescale of this diffusion depends on the orbit's perihelion (which determines the giant planet that is primarily delivering the kicks) and the current semimajor axis (which determines how often perihelion passages occur as well as the orbit's binding energy). As semimajor axis increases, the diffusing timescale for further changes in semimajor axis decreases with $a^{-1/2}$ for a fixed $q$. Meanwhile, a competing timescale is the typical time for a significant change in perihelion due to perturbations from passing stars and the Galactic tide. A torque large enough ($\Delta q\gtrsim10$ au) to potentially pull an orbit's perihelion outside of the planetary region will prevent future energy kicks from planetary encounters and halt diffusion in semimajor axis. This second timescale also decreases with semimajor axis since external perturbations become more significant as the distance from the Sun's gravitational pull increases, but it falls more steeply than the diffusion timescale, or as $a^{-2}$. 

\begin{figure}[htb]
    \centering
    \includegraphics[scale=0.54]{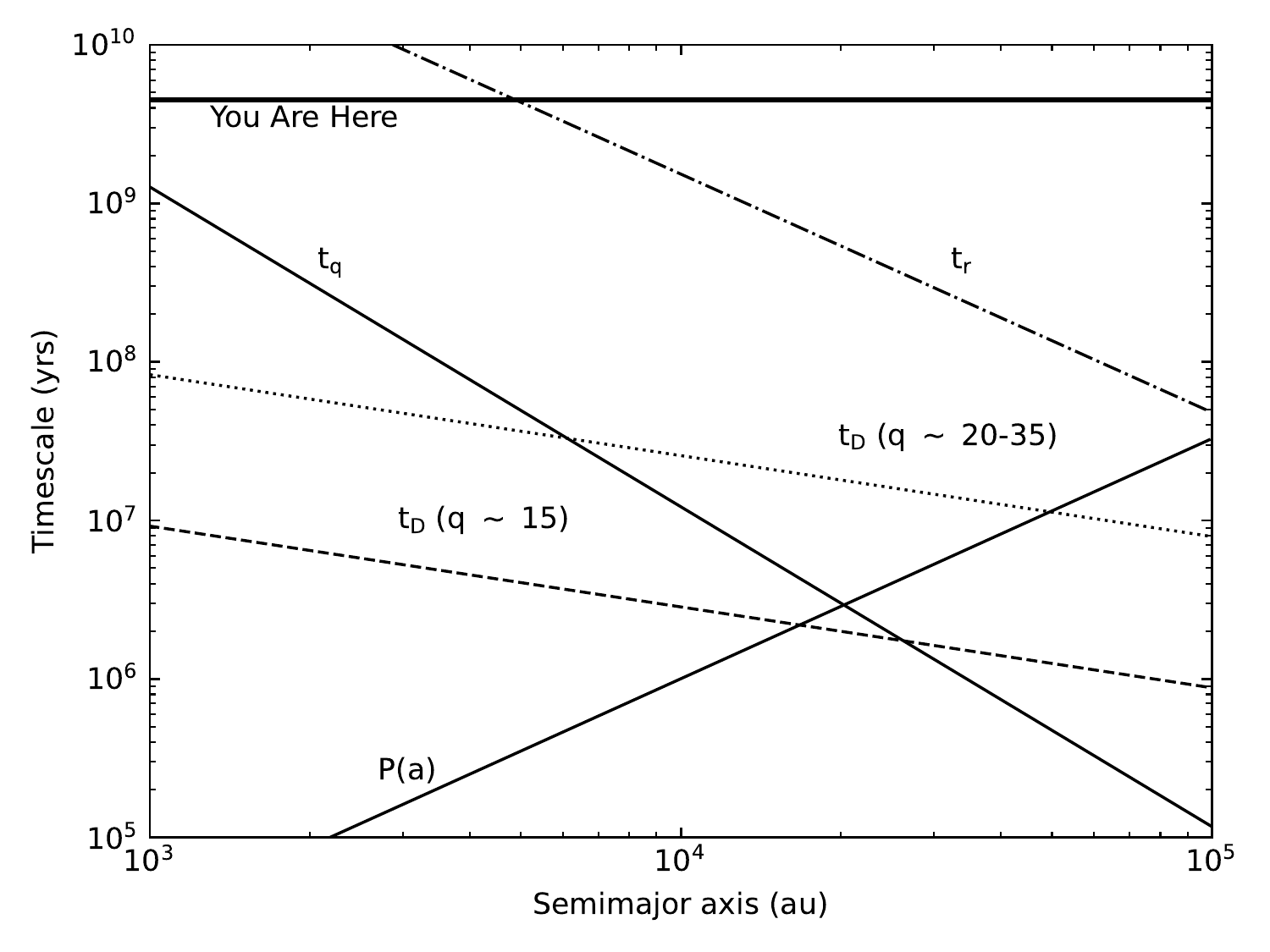}
    \caption{Plot of key dynamical timescales as a function of particle semimajor axis. $t_q$ is the timescale for a perihelion shift of 10 au (assuming initial $q$ of 25 au) due to Galactic tidal torquing; $t_D$ is the diffusion timescale for the semimajor axis to change by order of its present value due to planetary energy kicks; $t_r$ is the timescale for the Galactic tide to cycle an orbit from $q=25$ au to its maximum perihelion and back again; $P(a)$ is the orbital period; Horizontal line marks the age of the solar system. Adapted from \citet{dqt87}.}
    \label{fig:aq-timescales}
\end{figure}

Both the timescale for semimajor axis diffusion due to planetary encounters and the timescale for perihelion torques due to external perturbations are plotted as a function of semimajor axis in Figure~\ref{fig:aq-timescales}. 
To avoid a case where all planetesimals are either ejected or collide with the sun or planets, the timescale for a perihelion torquing must fall below the semimajor axis diffusion timescale at some value of semimajor axis. 
The semimajor axis for which the torquing timescale is shorter than the diffusion timescale depends on the planetesimal's initial perihelion distance because Jupiter and Saturn provide larger energy kicks than Uranus and Neptune.
For planetesimals being scattered by Uranus and Neptune, the two timescales become comparable at $a\simeq5000$ au, and for planetesimals being scattered by Jupiter and Saturn, they become comparable at $a\simeq20,000$ au. 
These semimajor axis values determine the inner edge (i.e., minimum $a$) of the Oort cloud, interior to which objects' perihelia cannot be changed by external perturbations before they experience significant changes in semimajor axis; objects interior to this $a$ boundary are part of the trans-Neptunian scattering population rather than the Oort cloud.
Exterior to this value, objects' perihelia can be lifted out of the planetary region, effectively locking their semimajor axes. 
Thus, bodies gravitationally scattering off the giant planets can enter the Oort cloud at or beyond the semimajor axis at which the two timescales become comparable. 
There is also a natural maximum semimajor axis to the Oort cloud that exists due to the tide of the Milky Way's disk, which overwhelms the Sun's gravitational pull around 0.5--1 pc or 1--2$\times10^5$~au \citep{heitre86}. 

With initial semimajor axes of 2000 au and bins of initial perihelia between 5 au and 35 au, \citet{dqt87} directly integrated $\sim$7000 particles under the influence of the Sun, planets, passing stars, and the Galactic tide for 4.5 Gyrs. At the end of this simulation, they found that a substantial fraction of particles were trapped in the Oort cloud with perihelia beyond the planets and semimajor axes between a few thousand au and $\sim$10$^5$ au. However, the trapping fraction varied greatly depending on particles' initial perihelia. For those with perihelia of 5 au (near Jupiter), the trapping fraction was only 2\%, while the fraction approached 40\% for those with perihelia near Neptune (25--35 au). The reason for this is that the magnitude of energy kicks particles receive during perihelion passage sensitively depends on perihelion value \citep{fern78}. 

These energy kicks were first numerically measured by \citet{fern81}, and the typical energy kick per perihelion passage for different orbits is shown in Figure~\ref{fig:dEvsq}.  Objects passing near Jupiter and Saturn receive energy kicks more than an order of magnitude larger than those passing near Uranus and Neptune.  As already stated, for objects with perihelia near Jupiter and Saturn, their perihelia cannot be pulled from the planetary region until they attain semimajor axis $\gtrsim$10$^4$ au. However, the entire orbital energy window of the Oort cloud ($10^4 \lesssim a \lesssim 10^5$ au) is smaller than the typical energy kick these objects receive on a single perihelion passage. Thus, most of them overshoot the Oort cloud and leave the solar system on hyperbolic orbits. Since the opposite is true for objects scattering off Uranus and Neptune (the weaker energy kicks of these planets are smaller than the Oort cloud's orbital energy window), Uranus and Neptune are much more efficient at placing bodies into the Oort cloud. Within the populations of particles trapped in the Oort cloud, \citet{dqt87} found that the same perturbations (passing stars and the Galactic tide) that were responsible for perihelion shifts were strong enough to complete isotropize orbital inclinations and eccentricities beyond a semimajor axis of $\sim$5000 au. 

Another prediction of the numerical simulations of \citet{dqt87} was that the Oort cloud is strongly centrally concentrated, with 80\% of bodies orbiting on semimajor axes between 3000 and 20000 au and the remaining 20\% of bodies orbiting beyond $a>20000$ au. This finding was interpreted to imply that only the outer 20\% of the Oort cloud was capable of supplying observed long-period comets. The reason for this is that the perihelion shift timescale plotted in Figure~\ref{fig:aq-timescales} is longer than the orbital period for semimajor axes $a\lesssim20000$ au. Thus, objects on semimajor axes below 20000 au must first make a perihelion passage near Jupiter and Saturn before they would be capable of making a perihelion passage near Earth. However, as shown in Figure~\ref{fig:dEvsq}, the energy kick delivered during this first perihelion passage will typically be powerful enough to either eject the particle on a hyperbolic orbit or decrease its semimajor axis to a level where its perihelion evolution is extremely slow. 

\subsubsection{Lower Central Concentration and Trapping Efficiency}

From the above logic it was thought that LPCs only provided constraints on the outer periphery of Oort cloud beyond 20,000 au, but the inner region could contain large numbers of bodies as well \citep{hills81}. \citet{hills81} speculated that this inner portion could be $\sim$100 times more populous, yet would only supply comets near Earth during ``comet showers'' coinciding with rare, powerful stellar encounters \citep{hills81, hut87}. The simulations of \citet{dqt87} suggested that these inner population estimates were far too high. 

    Based on the high Oort cloud trapping efficiencies of objects gravitationally scattering off Uranus and Neptune in \citet{dqt87}, it was presumed that tens of percent of leftover planetesimals may have been captured into the Oort cloud after the giant planets formed. However, \citet{fern97} noted that many objects beginning on orbits near Uranus and Neptune evolve to Jupiter- and Saturn-crossing orbits before being scattered to large semimajor axes \citep{fernip84} and that this process may lower the fraction of material that is trapped in the Oort Cloud \citep{fernip81}. Numerical work by \citet{dones04} confirmed this. \citet{dones04} found that the actual fraction of initial bodies that are still trapped in the modern Oort cloud was much more likely no more than $\sim$5\%, and subsequent works have pointed to even lower fractions of 2--3\% \citep{kaibquinn08, kaib11}. The reason for this difference from \citet{dqt87} was that \citet{dones04} initialized their particles on nearly circular orbits between the giant planets, whereas \citet{dqt87} had to begin their particles (due to computing limitations of the era) on highly eccentric orbits with semimajor axes of 2000 au. The choice of \citet{dqt87} effectively locked the perihelia of their particles unless the Galactic tide or passing stars altered them. On the other hand, the approach of \citet{dones04} allowed the perihelia of bodies to significantly evolve before they were scattered to eccentric orbits. 

\begin{figure}[htb]
    \centering
    \includegraphics[scale=0.6]{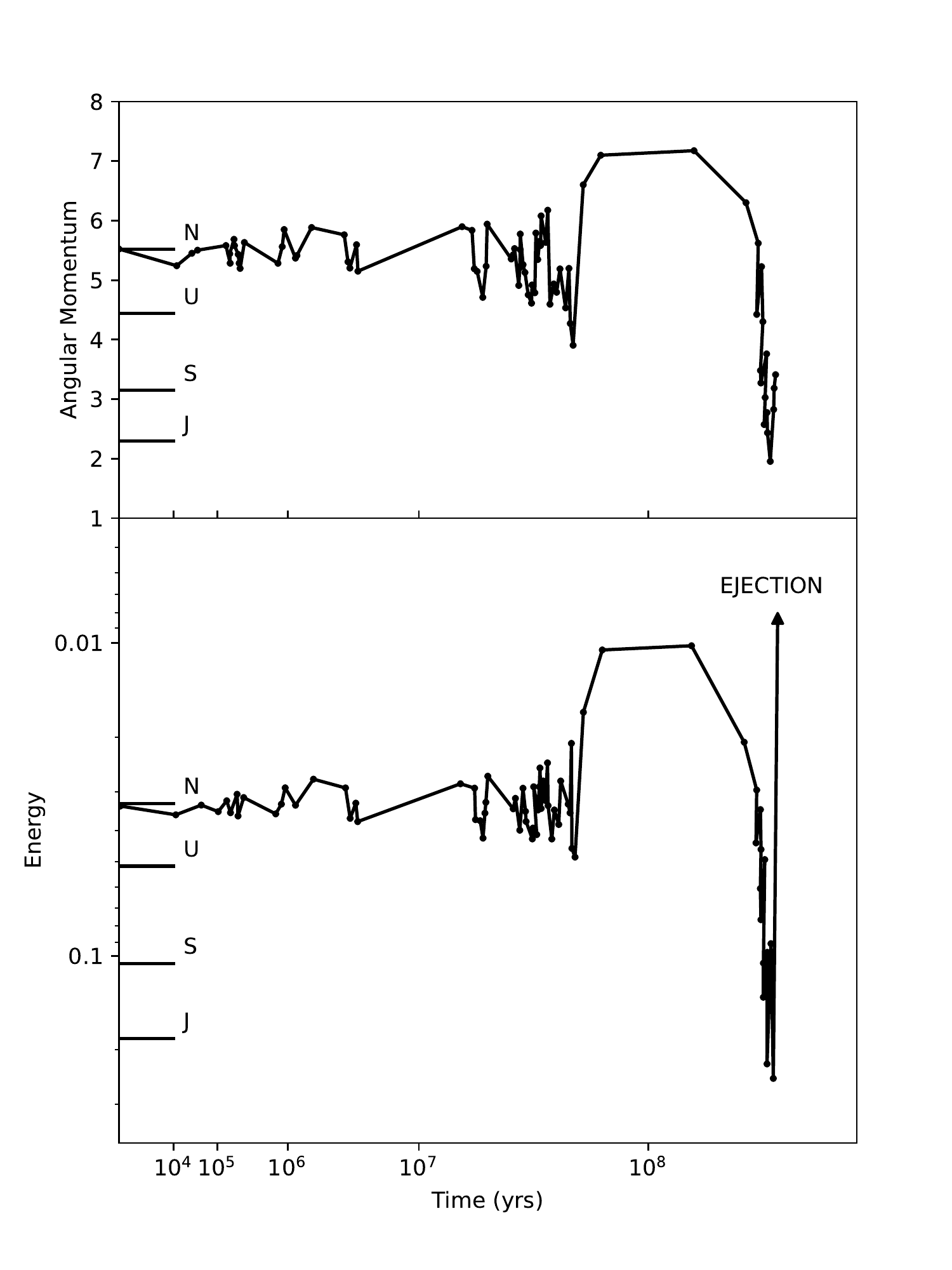}
    \caption{Evolution of a particle's angular momentum and orbital energy over time as it interacts with the giant planets. The energies and angular momenta of the giant planets are marked with letters along the $y$-axis. Energy and angular momentum units are such that $GM_{\odot}=1$ and distance is in au. Adapted from \citet{fernip84}.}
    \label{fig:fernip_qevol}
\end{figure}

Through earlier numerical experiments of particles scattering off of the giant planets done by \citet{fernip84}, it was found that particles beginning on circular orbits near the ice giants often first evolve to the Jupiter/Saturn region before being scattered to large semimajor axes. 
An example this evolution is shown in Figure \ref{fig:fernip_qevol}. 
This is the dynamical behavior that lowered the Oort Cloud trapping efficiency so much in \citet{dones04}. 
Although objects scattering off of Uranus or Neptune have a large probability of becoming trapped in the Oort cloud, most of the bodies that begin on circular orbits near these planets see their perihelia evolve toward Jupiter and Saturn, and, as discussed above, these planets are much more likely to eject objects than allow them to become trapped in the Oort cloud. 
The diminished role of Uranus and Neptune in placing objects into the Oort cloud also changes the cloud's structure. 
Since objects scattered by Jupiter and Saturn have little chance of becoming trapped in the Oort cloud until they attain semimajor axes above 10--20 thousand au, the fraction of bodies retained with $a<20,000$ au is nearly equal to that with $a>20,000$ au \citep{dones04}. 
Thus, the inner region of the cloud, thought to be unconstrained by observed long-period comets, was predicted to be comparably populous to the outer region beyond 20,000 au. 

\subsection{Generating a Scattered Disk} \label{sec:buildingSD}

The Oort cloud as described above is thought to be populated from an initially dynamically cold planetesimal disk in the current giant planet region.
However, the existence of the SPCs revealed the need for an extension of this disk past Neptune to provide a low-inclination source of these non-isotropic comets (see Section~\ref{ss:introSD}). It was then realized that the dynamical erosion of this disk that could feed the SPCs naturally leads to the formation of a scattered disk of objects with semimajor axes inside the inner Oort cloud boundary and perihelia near Neptune (\citealt{Torbett:1989,dunlev97}; though as discussed below much of the scattering population is now thought to originate from the initial planetesimal disk located interior to Neptune's current orbit).
Such objects were observational confirmed amongst the early TNO discoveries \citep{Luu:1997}. 


However, as more TNOs were discovered, the complex dynamical structures revealed in the classical, resonant, and scattering populations \citep[recently reviewed by][]{Gladman:2021} showed that the giant planets did not form on their current orbits. 
Early work on the interactions between the giant planets and the initially massive planetesimal disk showed the energy and angular momentum exchange can cause the planets' orbits to migrate \citep{fernip84}. 
Outward migration of Neptune was then shown to lead to capture of objects into its external mean motion resonances. 
The large eccentricity of Pluto's 3:2 resonant orbit was potentially explained by Neptune's outward migration by $\sim7$~au \citep{Malhotra:1993,Malhotra:1995}.
The large populations of objects in many of Neptune's resonances that have been found in subsequent surveys has provided strong evidence of migration-related resonant capture.





Our picture of the dynamical history of the giant planets has evolved significantly since the early proposals of so-called smooth planetary migration. 
To explain both the complex orbital distribution of the small bodies in the outer solar system and the dynamical state of the giant planets' orbits, more violent scenarios have been proposed.
The Nice model of the giant planets' evolution was one such early model; \cite{Tsiganis:2005} suggested that Jupiter and Saturn crossed a strong mutual mean motion resonance while migrating, triggering an epoch of large eccentricities in all four giant planets' orbits that led to planet-planet encounters and even a swap in the order of the ice giants.
This specific proposed resonance crossing has since been shown to not satisfactorily reproduce the orbits of Jupiter and Saturn \citep[e.g.,][]{bras09}, but the concept of excited planetary eccentricities, potential planet-planet gravitational encounters, or even ejection of an ice giant (see \citealt{Nesvorny:2018} for a recent review of the giant planets' dynamical history) hold promise for explaining some of the TNO orbital distribution features.



The current picture of the giant planets' dynamical history is that a mix of violent and smooth migration likely took place (see, e.g., \citealt{Nesvorny:2018}).
Their dynamical evolution dispersed the massive disk of planetesimals that formed in the current giant planet region into the trans-Neptunian and Oort cloud populations.
A small portion of the observed TNOs (the cold classical population on low-eccentricity, low-inclination orbits in the 42-48~au region) represent objects that formed in-situ beyond Neptune in a low-mass density disk (see, e.g., results from the New Horizons fly-by of the classical TNO Arrokoth; \citealt{McKinnon:2020}).
The majority of the TNO populations (the dynamically hot populations, including the scattering population) were scattered outward from  closer-in, more massive disk with an outer edge near 30 au. A small fraction of these scattered bodies survived in metastable orbits within the modern TNO population. This same scattering process also populated the Oort cloud.

\begin{figure}[ht]
    \centering
    \includegraphics[width=0.45\textwidth]{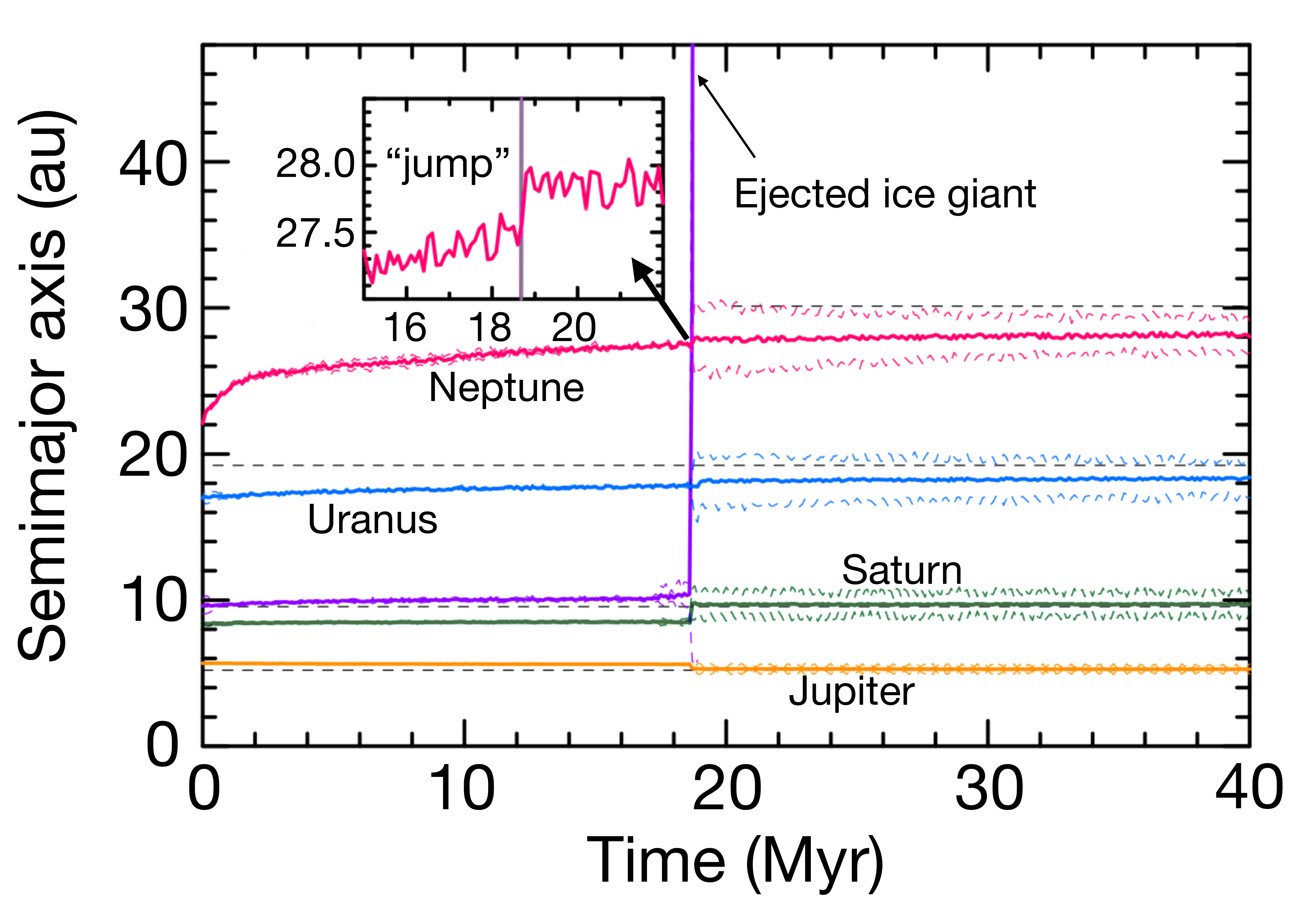}
    \caption{One plausible version of the giant planets' orbital histories from \cite{Nesvorny:2016} (modified from their Figure 3). Semimajor axis vs time for each planet is shown, with thinner dashed lines indicating the perihelion and aphelion distances of each planet. This migration scenario involves an additional ice giant that is ejected after $\sim18$~Myr, causing moderate eccentricity excitation and small jumps in the semimajor axes of several planets, including Neptune (shown in the inset). Neptune slowly migrates to its final location in the final $\sim100$~Myr of the simulation.}
    \label{fig:jumpingneptune}
\end{figure}

It is clear from modeling that smooth migration into an extended massive disk (one that extends out to $\sim50$~au) significantly over-populates Neptune's resonances compared to the observed populations \citep[e.g.,][]{Hahn:2005}.
The other motivation for an end to the massive initial planetesimal disk at $\sim$30~au is also, of course, that Neptune stopped migrating where it did (i.e., ran out of fuel for migration; \citealt{Gomes:2004}). 
But even assuming a transition to a less massive disk at $\sim30$~au, something besides smooth migration is needed to produce the detailed distribution of objects in the classical Kuiper belt as well as the distributions of objects in the scattering, detached, and distant resonant/near-resonant populations. 
The dynamical excitation from very violent scattering events (such as in the original Nice model) can still be consistent with retaining the cold classical population in certain circumstances \citep[e.g.,][]{Batygin:2010,dawmur12,Gomes:2018}, though more recent models investigated less dramatic instabilities.
Smaller scale `jumps' in Neptune's orbit caused, for example, by the ejection of an additional ice giant (see, e.g., \citealt{nes11,Nesvorny:2012}),  might help explain the `kernel' (an over-density of objects near 45~au noted by \citealt{Petit:2011}) in the cold classical region \citep{nes15b}; Figure~\ref{fig:jumpingneptune} shows an example of such an evolution for the giant planets.

In addition to planet-planet interactions, the number and mass of dwarf-planets in the planetesimal disk affects how `grainy' the planets' migration is; scattering events between 
\begin{figure}[t!]
    \centering
    \includegraphics[width=0.4\textwidth]{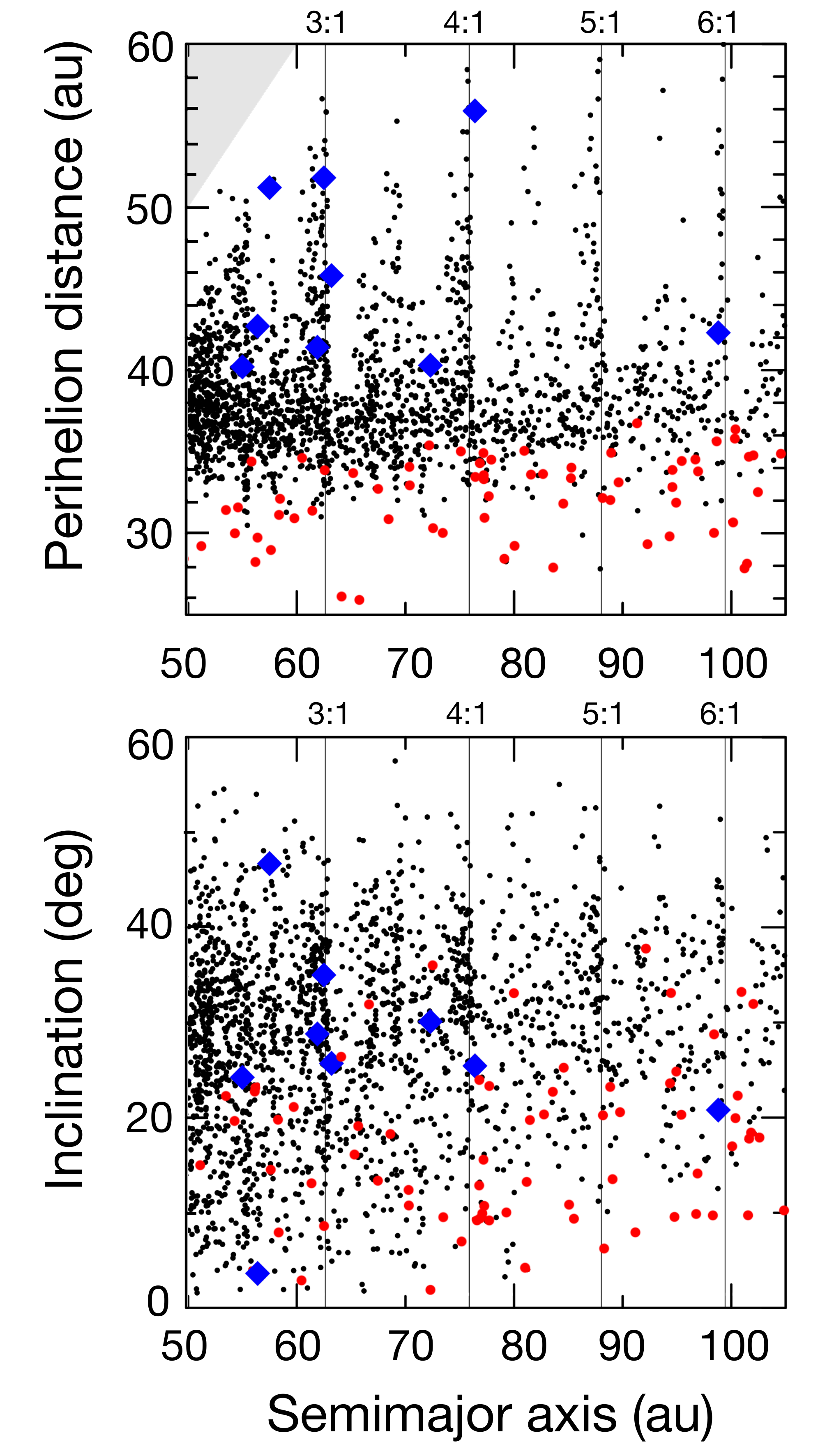}
    \caption{
    An example simulated distribution of the TNO populations from $a=50-100$~au from \cite{NVR16} (modified from their Figure 1) produced from a giant planet history scenario similar to that shown in Figure~\ref{fig:jumpingneptune} that also included grainy migration of the planets due to interactions with Pluto-mass objects.
    Perihelion distance (top) and inclination (bottom) vs semimajor axis are shown for the simulated particles that ended up in the scattering population (red dots) and the high-$a$ detached and resonant populations (black dots); a subset of the then-observed TNOs with $q>40$~au are shown (blue diamonds) for comparison.
    \label{fig:migoutcome}
    }
\end{figure}
Neptune and these largest objects in the disk cause small but significant jumps in Neptune's semimajor axis that can alter the capture rate into Neptune's resonances  \citep[e.g,][]{Murray-Clay:2005,Nesvorny:2016}.
The speed of migration, likely determined by the mass distribution of the planetesimal disk, also affects the expected TNO distribution.
There are models that predict that some period of slow migration, particularly during the last stage of migration after any planet-planet scattering has occurred, can help explain the high-perihelion objects in the scattering/detached TNO population by allowing for  secular-resonant interactions to occur, which might also help excite the inclinations of the hot populations  \citep[e.g.][]{nes15a,Lawler:2019,Fraser:2017}.
Adding graininess to this slow migration can help strand objects out of the scattering/resonant population by dropping them out of resonances at a high-$q$ portion of this secular evolution, building up a population of near-resonant detached objects \citep[e.g.,][]{kaibshep16, Lawler:2019}.
Figure~\ref{fig:migoutcome} shows an example of the scattering/detached/resonant populations from $a\simeq50-100$~au that are expected for a planetary migration scenario with a small instability followed by a long period of grainy migration (from \citealt{Nesvorny:2016}).
Observations of the high-$a$, high-$q$ TNOs remain relatively sparse (as indicated by the relatively small number of observed objects included in Figure~\ref{fig:migoutcome}) due to the strong observational biases against detecting them, but future surveys will help provide more robust observational constraints on models of the scattering and detached populations (see Section~\ref{sec:future}).

\subsection{Concurrent Oort Cloud and Kuiper Belt Formation}

Over the past 3 decades, evidence has mounted that planetesimal scattering drove the giant planets through substantial orbital evolution early in the solar system's history. However, Oort cloud formation, a product of this scattering, was not simulated in concert with giant planet orbital evolution until the work of \citet{brasmorb13}. In this work, the authors replicated the post-instability behavior of the surviving giant planets within an early iteration of the Nice Model \citep{Levison:2008}. Neptune and Uranus began with eccentricities above 0.2 and damped to their modern values while their semimajor axes migrated outward 2--3 au over the course of tens of Myrs. At the initiation of this sequence, test particles were started with semimajor axes between 29--34 au, eccentricities of 0.15, and coplanar inclinations. It was this outer belt from which \citet{brasmorb13} assumed the Oort cloud (and scattered disk) is formed. 

After 4 Gyrs of evolution \citet{brasmorb13} found an Oort cloud trapping efficiency of $\sim$7\%, modestly higher than \citet{dones04} and more than double that of \citet{kaibquinn08}. While giant planet orbital evolution may have contributed to this, the confinement of test particles to beyond $a\geq29$ au may also have been responsible. No particles were begun in the immediate vicinity of Jupiter and Saturn, which are very inefficient Oort cloud populators, while the contributions of Uranus and Neptune were maximized. 

In addition to forming the Oort cloud, the simulations of \citet{brasmorb13} also generated a scattered disk. The population of this scattered disk was $\sim$12 times smaller than that of the Oort cloud. This result was in line with the $\sim$10:1 population ratios found in \citet{dones04} and \citet{kaibquinn08}. Meanwhile, comparisons of LPC flux with the steady-state populations of JFCs have historically been interpreted as implying a $\sim$100:1, or perhaps even higher ratio \citep[see \ref{sec: ratio} as well as][]{lev10}. 

The planetary orbital evolution of \citet{brasmorb13} utilized an early, violent version of the Nice model that was demonstrated to conflict with numerous aspects of the modern  solar system's architecture \citep{bras09, bat11, dawmur12}. As discussed in Section~\ref{sec:buildingSD}, recent detailed comparisons of TNO surveys with simulations of Kuiper belt formation appear to favor a less violent giant planet evolution in which Neptune's migration was relatively smooth from 24 to 30 au except for a brief interruption during the ejection of an additional ice giant \citep{nes11, nes15a, nes15b}. Incorporating this later iteration of early giant planet evolution, \citet{vok19} again modeled the formation of the Oort cloud. In this instance, particles were initiated on circular orbits between 24--30 au while Uranus and Neptune migrated several au to their modern semimajor axes over $\sim$100 Myr timescales. Although this work focused on modeling the comet production from such a cloud, the authors found that $\sim$6\% of their bodies were trapped in the Oort cloud after 4.5 Gyrs of evolution, in line with the results of \citet{brasmorb13}, likely due to the backweighted distribution of initial test particle orbits. In addition, \citet{vok19} found that the inner Oort cloud has nearly the same population size as the outer Oort cloud, although they divided the inner and outer clouds at $a=15,000$ au. 

\begin{figure}[htb]
    \centering
    \includegraphics[scale=0.6]{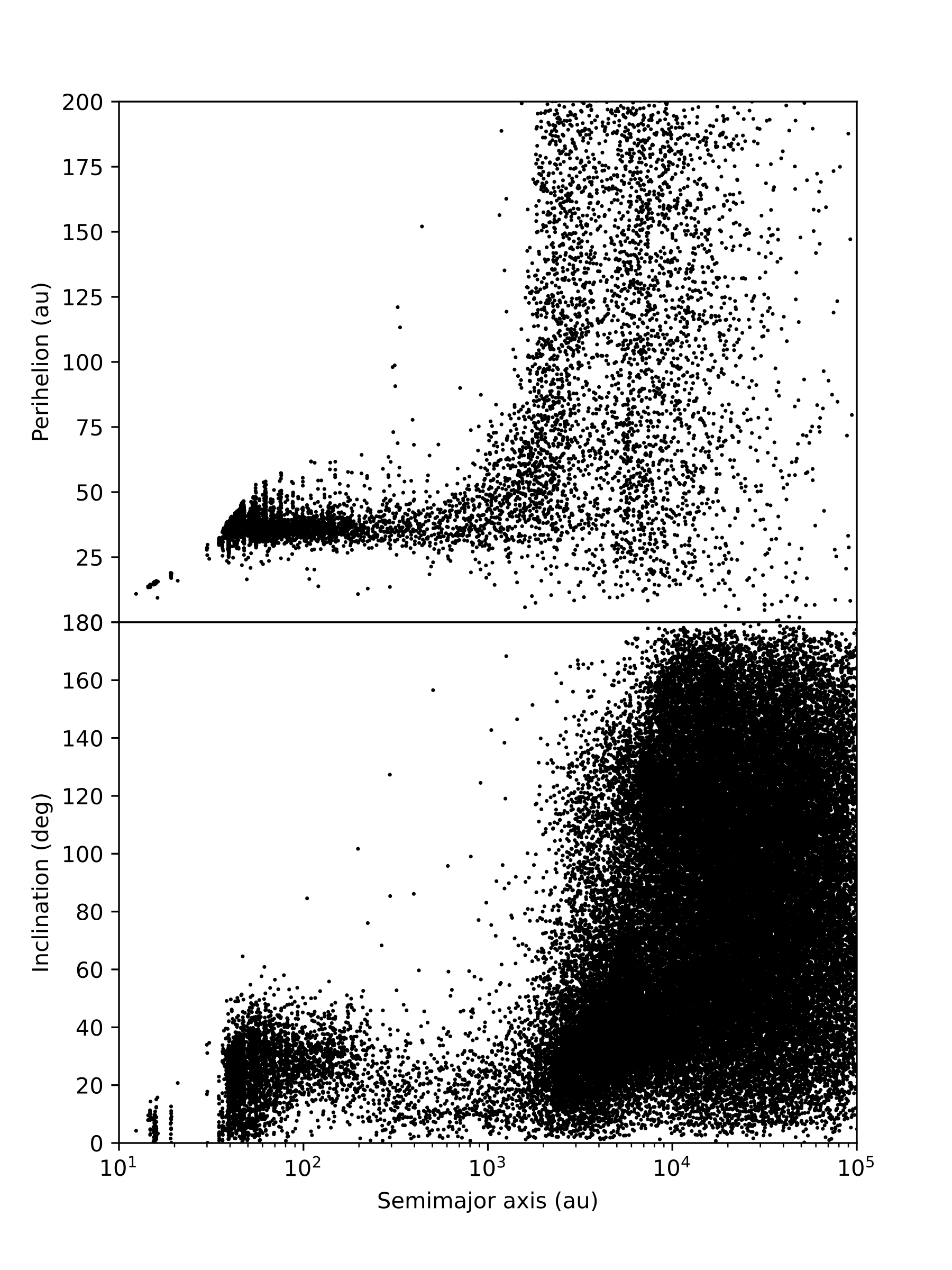}
    \caption{Perihelia ({\it upper panel}) and inclinations ({\it lower  panel}) vs semimajor axes of particles at $t=4$ Gyrs in a simulation forming the Oort cloud and scattered disk concurrently. Plot is generated from the ``OC'' simulation of \citet{kaib19}.}
    \label{fig:kaib19}
\end{figure}

For illustrative purposes, Figure \ref{fig:kaib19} shows the distribution of particle semimajor axes, eccentricities, and inclinations after 4 Gyrs of evolution in an Oort cloud formation simulation. This simulation, taken from \citet{kaib19}, forms an Oort cloud and scattered disk concurrently with very similar parameters to the work of \citet{vok19}. As one can see, the scattered disk generally has perihelia between $\sim$30 au and $\sim$45 au and inclinations below $\sim$45$^{\circ}$ until semimajor axes above $\sim$10$^3$ au. At this point, perturbations from the local Galaxy begin to isotropize inclinations and detach perihelia from the planetary region.

\subsection{Influence of the Sun’s Local Environment}

\subsubsection{Solar Birth Cluster}

While the structure of the scattered disk should be relatively insensitive to the Sun's local galactic environment, perturbations from this environment are critical to the formation and evolution of the Oort Cloud, and variations in this environment could impact the Oort Cloud's formation. The numerical work of \citet{dqt87} and \citet{dones04} assumed that the Sun's local galactic environment did not significantly change with time. The strength of the Galactic tide was based on the local density of matter in the Milky Way disk inferred from stellar motion \citep{bahcall84}, and the velocities and rates of passing stars were meant to mimic the population and kinematics of the solar neighborhood \citep[e.g.][]{garcia01, bail18}.

This static galactic environment was, of course, an approximation, and there are reasons to expect it to have changed significantly over the Sun's lifetime. Perhaps most significantly, it is well-known that most stars form within clusters and spend some portion of their early lives within them \citep{lala03}. 
If this occurred for the Sun, the local stellar densities during this cluster phase would have been orders of magnitude higher than the modern solar neighborhood, and the tide of the cluster would have dwarfed the current Galactic tide perturbing the modern solar system. 
\citet{gaidos95} realized that these enhanced external perturbations could have dramatic impacts on the formation of the Oort cloud, allowing objects to become trapped within the Oort cloud at a much lower semimajor axis than predicted for a galactic field environment. However, \citet{gaidos95} argued that such a comet cloud would also be rapidly destroyed by the same cluster perturbations and that little material should remain at these distances.

\citet{fern97} further considered the scenario in which Oort cloud formation begins within a stellar birth cluster. He noted that the heightened perturbations of a cluster environment would dramatically widen the energy window of the Oort cloud and potentially allow Jupiter and Saturn to play more prominent roles in populating the cloud. Moreover, much of this material could reside at hundreds of au to a few thousand au, well interior to the semimajor axis range of the ``classical'' Oort cloud. Moreover, \citet{fern97} argued that the relatively short lifetimes of open and embedded clusters would have allowed many of these bodies to remain trapped, contrary to the conclusions of \citet{gaidos95}. 

The formation of the Oort cloud within a star cluster environment was then first simulated in \citet{fernbrun00}. In this work, particles were initiated on semimajor axes between 100--250 au with perihelia distributed amongst the giant planets. The entire solar system was then immersed within a star cluster environment that contained stars passing much more slowly (1 km/s) and closely than a Galactic field environment. The stellar densities of clusters were varied between 10 and 100 stars/pc$^3$ and linearly declined to zero over 10$^8$ years. In addition, \citet{fernbrun00} included tidal torquing from a uniform-density spherical distribution of gas that dispersed at 10$^7$ years meant to replicate the gaseous component of an embedded cluster. They found that an inner comet cloud readily formed under such conditions and the radial extent (semimajor axes from a couple hundred au to a few thousand au) depended on the stellar density of the star cluster, with denser clusters yielding more compact clouds. Moreover, as in \citet{dqt87}, the fraction of material captured into this cloud depended sensitively on particles' initial perihelion. Those with perihelia in the Jupiter/Saturn region only had trapping efficiencies of a few percent, whereas tens of percent of material could be captured into the cloud for initial perihelia near the ice giants. Finally, the relatively short cluster lifetimes allowed these inner clouds to largely stay intact past cluster dissolution. 

\begin{figure}[htb]
    \centering
    \includegraphics[scale=0.53]{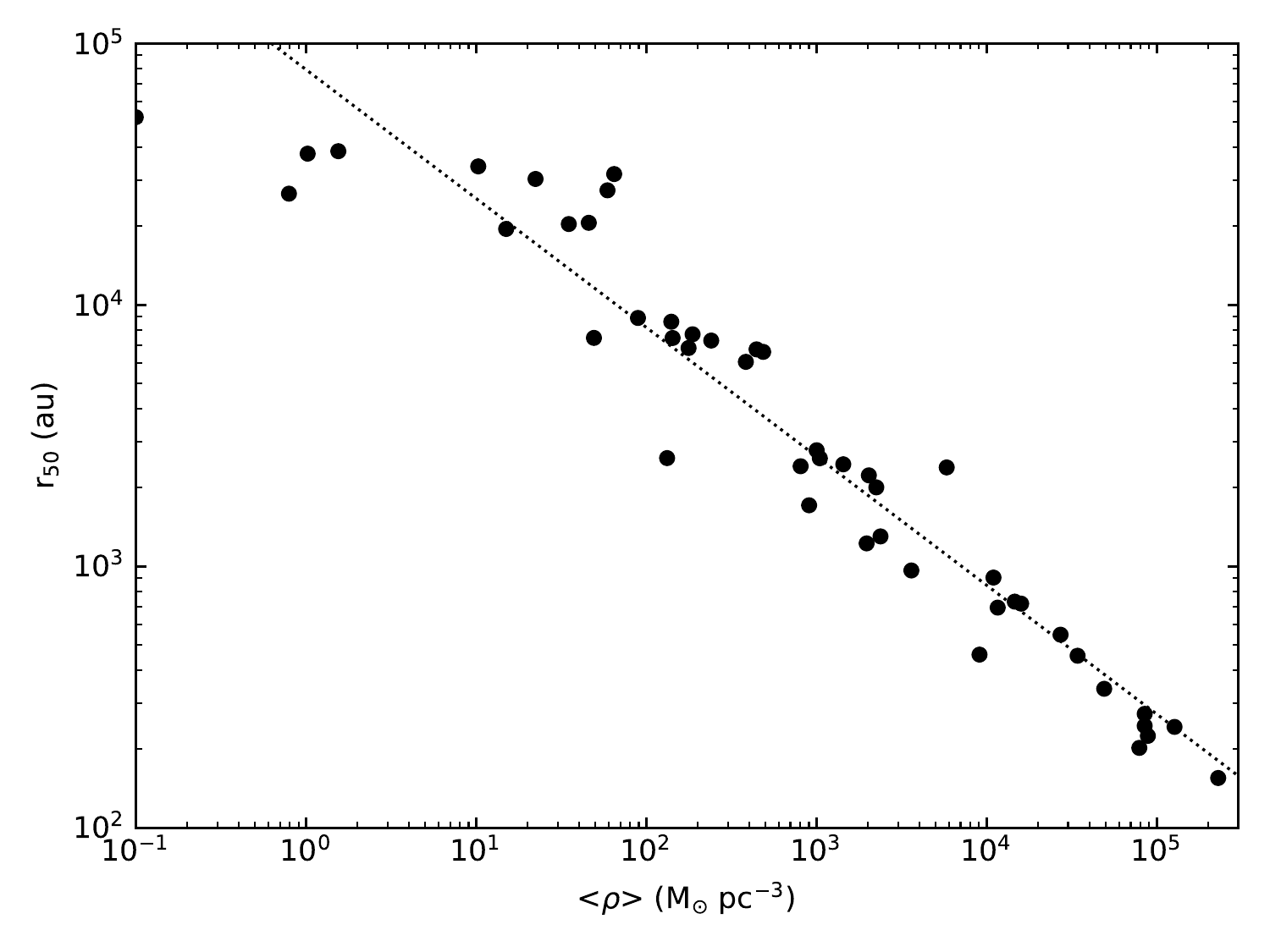}
    \caption{Median position of bodies within the Oort cloud as a function of the mean density of a cluster environment for a given Oort cloud formation simulation. From \citet{Brasser:2006}.}
    \label{fig:bras06}
\end{figure}

Oort cloud formation within a cluster was revisited by \citet{Brasser:2006}. This work specifically focused on the embedded cluster environment, in which most stars appear to form \citep{lala03}. The embedded cluster environments employed within \citet{Brasser:2006} had higher central densities (in gas and stars) and shorter lifetimes than those used in \citet{fernbrun00}. Due to the large energy window of the comet cloud and the short cluster lifetimes, roughly 10\% of material scattered by Jupiter and Saturn could be caught in an inner comet cloud, whose range of semimajor axes again depended on the assumed density of the cluster environment (see Figure \ref{fig:bras06}). Nearly all of this inner comet cloud material would be retained until the present epoch and very little would be lost or diffuse to the classical Oort cloud beyond $a>20,000$ au \citep{bras08}. 

Additional investigations by \citet{bras07} studied the effect of the gaseous component of the Solar Nebula during the formation of the Oort cloud within a star cluster. For small km-sized bodies, aerodynamic drag prevented objects from being scattered to large semimajor axes. Instead, they concentrated into nearly circular orbits interior to Jupiter. Meanwhile, objects with radii of tens of km were decoupled from the gas and freely reached the Oort cloud. If the gaseous component of the Sun's protoplanetary disk outlived the Sun's birth cluster, this suggested that the centrally concentrated, cluster-generated comet cloud should only contain large ($r\gtrsim10$ km) bodies. 

\citet{Brasser:2012} repeated the embedded cluster scenario with improved simulations that used updated, observationally motivated cluster profiles and stellar velocity distributions. Furthermore, in this new work cluster stars were able to gravitationally interact with one another. While centrally concentrated Oort clouds still formed down to semimajor axes of several hundred au, the formation efficiency dropped to 1--2\% due to the Oort cloud harassment that occurred with a heightened number of stellar interactions in these new environments. 

Complementing studies of Oort cloud formation within embedded clusters, \citet{kaibquinn08} simulated the formation of the Oort cloud within longer-lived ($t\simeq10^8$ years) open cluster environments. Unlike the work of \citet{fernbrun00}, particles were initiated on circular orbits. As expected, this allowed particles near Uranus and Neptune to evolve to Jupiter- and Saturn-crossing orbits before being scattered to large semimajor axes. This led to lower ($<10$\%) trapping efficiencies than predicted by \citet{fernbrun00}. Owing to the clusters' longer lifetimes and the more prominent role of powerful stellar encounters compared to the gas-rich environments of \citet{Brasser:2006}, the Oort clouds formed within open clusters displayed lower overall trapping efficiencies ($\lesssim5$\%) and a greater level of stochasticity in the degree of central cloud concentration (which was largely dependent on the few most powerful stellar encounters). Nonetheless, this work showed that an Oort cloud formed within an open cluster should possess bodies on semimajor axes from several hundred to several thousand au whose numbers are comparable to those in the classical Oort cloud. 

\subsubsubsection{Discovered Objects as Possible Birth Cluster Relics}

Beginning with the pioneering works of \citet{gaidos95} and \citet{fern97}, it has been predicted that if the Sun formed within a star cluster, we should expect the inner edge of the Oort cloud to potentially reside at a semimajor axis of hundreds of au rather than the thousands of au suggested by models that only consider the modern solar neighborhood. These predictions seem extremely prescient, as in the early 2000's TNO surveys began discovering objects with semimajor axes of hundreds of au but perihelia that dynamically decoupled them from planetary energy kicks \citep{glad02}. Although some of these bodies may attain their high perihelia via Kozai-Lidov cycles within Neptunian mean motion resonances, this mechanism cannot generate all observed orbits \citep{brassschwamb15, gom05, koz62, lid62}. In particular, the objects Sedna and 2012 VP$_{113}$ both have perihelia beyond 70 au and semimajor axes beyond 250 au and are therefore not susceptible to Kozai-Lidov cycling \citep{brown04, trushep14}. 

The discoveries of such objects appear consistent with the existence of a cluster-generated Oort cloud \citep{morblev04}. However, other explanations have been put forth, such as scattering off of an $\sim$Earth-mass planetesimal or exchange with another young star system \citep{gladchan06, kenbrom04, siltre18}. Of particular note, it has been posited that these detached TNOs with large semimajor axes are the dynamical signature of a distant, yet-undetected planet \citep{batbrown16a}. Such a planet has been shown to be capable of aligning the poles and perihelion directions of an ensemble of such orbits, and the detected orbits appear to display an anomalous anisotropy \citep{batbrown16a}.  However, the total sample size of objects remains small, and the observing biases associated with combining the results of many TNO surveys are complex \citep{shank17, kaib19}.
Detections from well-characterized surveys do not find evidence of significant orbital clustering \citep{shank17,Napier:2021,Bernardinelli:2021}.

\subsubsubsection{Oort Cloud Capture within a Birth Cluster}

Another aspect of Oort cloud formation within a cluster environment concerns the possibility of the Sun capturing exoplanetesimals, or bodies formed around another star cluster member and subsequently ejected into intracluster space. This possibility was first considered in \citet{eggers97} and then again in \citet{zheng90}. Both works found that forming our entire Oort cloud through the capture of exoplanetesimals within a cluster is unlikley. Such a scenario would require 10$^{16}$ planetesimals per star, which would be difficult to reconcile with the (at the time) non-detection of interstellar comets passing through the solar system \citep{zheng90}. 

\begin{figure}[htb]
    \centering
    \includegraphics[scale=0.38]{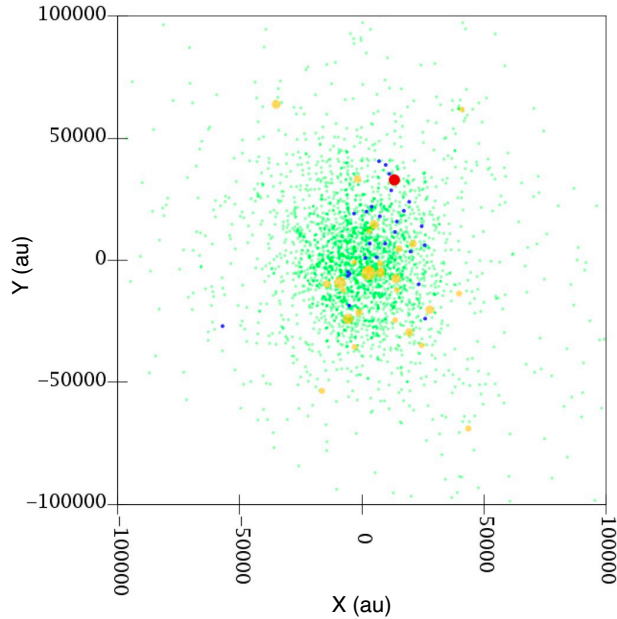}
    \caption{Snapshot of a simulation of Oort cloud capture within an embedded cluster. Datapoints mark the Cartesian X and Y positions of simulation particles. Yellow and red points are cluster stars, while green and blue points mark planetesimals. Blue planetesimals become bound to the red cluster star after cluster dissolution. From \citet{lev10}.}
    \label{fig:lev10}
\end{figure}

However, the capture probabilities used in those early works may have been underestimates. \citet{lev10} directly simulated the dynamical evolution and dissolution of embedded clusters in which each star was surrounded by a disk of distant planetesimals. Over time, many planetesimals became unbound from their stars-of-origin via cluster tides and star-star encounters. A snapshot of one simulation is shown in Figure \ref{fig:lev10}. In these simulations, it was found that stars often captured other stars' planetesimals. As stars and intracluster planetesimals escaped the potential of the dissolving cluster, some star and planetesimal velocity vectors were inevitably similar. The escape from the potential decreased relative velocities even further, enabling capture. In this scenario, stars typically captured a couple percent of their original planetesimal number through this mechanism. Thus, the efficiency of such capture should generally not exceed the efficiency of the solar system capturing its own scattered material through the ``classic'' Oort cloud formation process. However, occasionally direct exchange of planetesimals through a close star-star encounter led to capture fractions over 10\%. From these results, \citet{lev10} argued that a significant fraction, and perhaps the majority, of bodies in the Oort cloud could have originated around other stars. However, many of these bodies were captured onto distant orbits with semimajor axes of tens of thousands of au, and their subsequent evolution and survival within the Galaxy until the present epoch was not modeled. 

\subsubsection{Galactic Migration}

Even after the Sun's birth cluster disperses, there is ample reason to expect significant changes in the Sun's local galactic environment. Dynamical simulations of spiral galaxies indicate that the galactocentric orbits of most stars are not static with time. Instead, they often exchange angular momentum with the spiral arms, typically within corotation resonances, whose locations fluctuate as spiral wave pattern speed varies \citep{sellbin02, rosk08}. On average, this transports disk stars outward over time (although inward transport is also possible). Importantly, this process does not necessarily increase the peculiar velocities of stars, and they can remain on nearly circular orbits about the galactic center with modest oscillations above and below the galactic midplane. 

This stellar migration process obviously has large ramifications for our understanding of spiral galaxies and their properties, but it has also raised the prospect that the Sun may have undergone significant migration (likely outward) during its lifetime in the Milky Way. If this were the case, this may have significant implications for the formation of our Oort cloud. As the Sun's galactocentric distance changes, the mean density of its local environment will also change, which will modulate the rate of stellar encounters and the strength of the Galactic tide \citep{mat95}. \citet{bras10} conducted simulations of Oort cloud formation at various galactocentric distances of 2, 4, 6, 8, 14, and 20 kpc using an analytic model for the Milky Way potential. Unsurprisingly, the different environmental densities varied the Oort cloud inner edge (which was defined as the 5th percentile in semimajor axis) from 30,000 au at $r=20$ kpc to 2,000 au at $r=2$ kpc. In addition, the Oort cloud trapping efficiency varied between 2--4\% and did not display a strong dependence on galactocentric distance. 

However, the simulations of \citet{bras10} still assumed a static (local and global) galactic environment and tide for the each entire simulation. In fact, stellar migration within galaxies implies that the local environment and tide should be changing with time. The work of \citet{kaib11} addressed this by merging Oort cloud formation simulations with data output from the evolution of a Milky Way analog spiral galaxy \citep{rosk08}. Within the galactic simulation, an ensemble of 31 star particles were selected that possessed Sun-like galactic kinematics and ages. The histories of these particles were then backtraced throughout the simulation. These particles displayed a wide variety of behaviors. On average, their galactocentric distance varied by $\sim$5 kpc throughout their lifetimes, and the closest approach to and furthest excursion from the Galactic center was 2 kpc and 13 kpc, respectively. At each galactic simulation time output, the local density and galactic tidal field about each star particle was measured, and with this data, a time-varying set of external perturbations (galactic tide and passing field stars) was built for each star particle, or for each example of a galactic dynamical history of a Sun-like star. These perturbation sets were then used to model the formation of the Oort cloud for 31 hypothetical dynamical histories within the Milky Way. 

\citet{kaib11} found that the galactic dynamical histories of their star particles had a significant impact on Oort cloud formation, with the median semimajor axis of the Oort cloud varying by nearly an order of magnitude across the sample of star particles. Generally, the minimum galactocentric distance attained, which typically determined the maximum local density of matter inhabited, had the largest impact on the Oort cloud structure. Excursions toward the Galactic center resulted in more centrally concentrated clouds. Overall, Oort cloud formation models that do not account for the Sun's dynamical history within the Galaxy likely overestimate the cloud's inner edge and median semimajor axis. 

In addition, the time at which a star attains its minimum galactocentric distance affected the Oort cloud trapping efficiency. If the minimum galactocentric distance was attained early in the system's life, the trapping efficiency was enhanced (to 3--4\%), because the Oort cloud energy window was widened early in the system's history, and it was subjected to somewhat weaker galactic perturbations later in time. 
Conversely, late excursions to small galactocentric distances decreased Oort cloud trapping efficiency to 1--2\%, as the energy window of the Oort cloud remained small early in the system's history, and the late inward excursion drove the loss of weakly bound bodies. 

\subsubsection{Critical Roles of Stellar Passages}

\begin{figure}[htb]
    \centering
    \includegraphics[scale=.54]{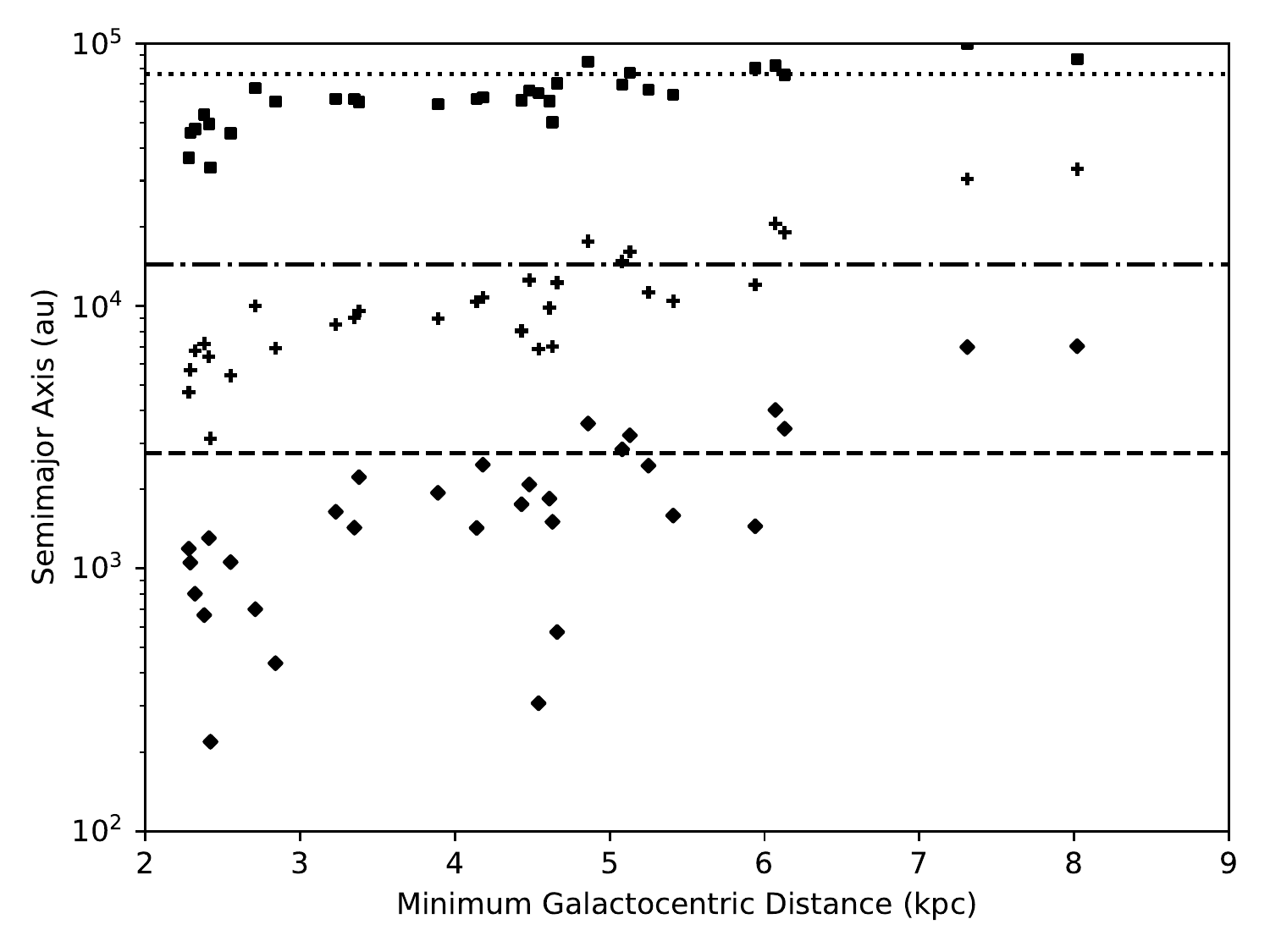}
    \caption{Plot of maximum ({\it squares}, 95th percentile), median ({\it crosses}, 50th percentile), and minimum ({\it diamonds} 5th percentile) semimajor axis of Oort cloud as a function of the minimum galactocentric distance attained by the solar analog within a galactic N-body simulation. Adapted from \citet{kaib11}.}
    \label{fig:KRQ11}
\end{figure}

As illustrated in Figure \ref{fig:KRQ11}, \citet{kaib11} also revealed that Oort cloud structure contains significant variation even after the Sun's potential migration history is considered, especially with respect to the Oort cloud's inner edge. The main source of this scatter is due to the effects of stellar passages. Since \citet{heitre86} showed the galactic tide to be a more powerful mean perturbing force than stellar passages, works studying the formation of the Oort cloud have often focused more on Galactic tidal effects and less on stars. However, the conclusions of \citet{heitre86} are only true {\it on average}. Passages of field stars are inherently stochastic events that occur on very short ($<10^5$ years) timescales, so there can be great variability in their significance from one epoch to another. Therefore, over the course of 4.5 Gyrs, there must have been short periods where the perturbation from stellar passages on the Oort cloud greatly exceeded the perturbation from the Galactic tide. Because the time to be scattered into the Oort cloud is much shorter than the age of the solar system, these brief, stellar-dominated periods of perturbation are ultimately what sets the very minimum semimajor axis at which the perihelia of bodies are torqued into the Oort cloud. 

When accounting for the role of stellar passages setting the Oort cloud's inner edge, there is a large amount of stochasticity since it depends on the few most powerful stellar passages over the Sun's history. To estimate the Oort cloud's very inner edge, \citet{kaib11} examined orbits that had only been weakly torqued from the planets ($60<q<100$ au) and measured the tenth percentile of semimajor axis. With this parameterization, they found that the Oort cloud's inner edge can vary by a factor of several (even with a fixed local galactic environment) due to the stochastic nature of stellar passages. In fact, this inner edge was pushed interior to Sedna's orbit in $\sim$20\% of their simulations, raising the prospect of whether a solar birth cluster (or other alternative mechanism) is necessary to explain this orbit. These effects of the most powerful stellar encounters can also be seen in the smattering of low-semimajor-axis particles with detached perihelia in Figure \ref{fig:kaib19}. An additional analysis of the stellar encounters experienced by the Sun for different galactic migration scenarios largely agreed with these results \citep{mart17}. 

Stellar encounters also play a critical role in maintaining the Oort cloud's orbital isotropy. The vertical term of the Galactic tide causes the orbital elements of Oort cloud bodies to fluctuate in a very regular manner \citep{heitre86}. Consequently, left unperturbed, the Oort cloud would exhaust the set of orbits that can attain very low ($q<15$ au) perihelion under the action of the Galactic tide, since these bodies are generally lost when they interact with Jupiter and Saturn. Before these regions of orbital space are completely depleted by the Galactic tide, however, the randomizing perturbations of stellar encounters replenish them \citep{rick08}. Thus, although the Galactic tide delivers most Oort cloud bodies into the planetary region as LPCs, stellar encounters are necessary to ensure that an abundant supply of Oort cloud bodies remain within the phase space that yields LPCs. 

\subsubsubsection{Recent Stellar Encounters with the Sun}

As our census of the Sun's nearby stellar neighbors has improved, numerous works have searched for particularly close stellar passages in the recent past as well as the short-term future ($\lesssim\pm$10 Myrs). Using Hipparcos data, \citet{garcia01} found several potential encounters within 1 pc of the Sun within $\pm$10 Myrs. In particular, they noted that the K star Gliese 710 should approach the Sun within 0.4 pc roughly 1.8 Myrs from now. None of the anticipated encounters were projected to deliver significant perturbations to the Oort cloud, but \citet{rick12} noted that the Hipparcos catalog was perhaps only $\sim$10--20\% complete over this timeline and uncertainties within astrometry and radial velocity prevented confident periastron calculations for close stellar encounters \citep{garcia01}.

The ongoing Gaia mission is assembling the three-dimensional space velocities of over 1 billion stars, four orders of magnitude greater than Hipparcos, and its astrometric precision will be two orders of magnitude finer than Hipparcos. \citet{rick12} concluded that the final data release of Gaia (anticipated in 2022) should reveal $\sim$90\% of the stellar encounters that have contributed to the flux of currently observable long-period comets. With the second release of Gaia data, \citet{bail18} found that $\sim$20 stars per Myr approach the Sun within at least 1 pc. In addition, with the improved astrometry of Gaia, they found that Gliese 710 has a 95\% probability of coming within 17,000 au of the Sun, and later refinements of this star's kinematics showed that it will come within 10$^4$ au 1.3 Myrs from now \citep{delafu20}. The stellar encounter flux of \citet{bail18} implies that such a close encounter should only be expected every $\sim$10 Myrs, and this is the closest known stellar encounter with the Sun. The improved stellar radial velocities coming with the third Gaia data release may still reveal other close passing stars in the near future or recent past. 

\subsubsection{Comet Showers}

Another well-studied effect that stellar encounters have on the Oort cloud dynamics are comet showers. Typically, the only bodies in the Oort cloud that can undergo a perihelion shift from beyond Saturn's orbit to near Earth's orbit within a single revolution about the Sun have semimajor axes $\gtrsim20,000$ au \citep{heitre86}. However, as described in the last section, the perturbative strength of field star encounters varies wildly over time. During the most powerful stellar passages that have occurred in the solar system's history, the perturbations delivered on the Oort cloud are temporarily strong enough to enable this perihelion shift for virtually any semimajor axis in the Oort cloud. This led \citet{hills81} to argue that most of the time, the Earth is only exposed to LPCs from the outer periphery of the Oort cloud, but during powerful stellar passages, the Earth is temporarily exposed to the entire Oort cloud, meaning that the flux of LPCs and potential Earth impactors could increase greatly during these events, which are called comet showers. 

The potency of comet showers clearly depends on how well-populated the inner ($a\lesssim$ 20,000 au) portion of the Oort cloud is relative to the outer cloud. \citet{hills81} assumed the inner Oort cloud could hold up to 100 times as many bodies as the outer cloud, and under this assumption, the most powerful stellar encounter of the past 500 Myrs would temporarily increase the flux of comets near Earth by $\sim$3 orders of magnitude. \citet{hut87} proposed that the impacts associated with such events could be responsible for one or more of Earth's mass extinctions. However, subsequent Oort cloud formation models predicted a much more modest inner Oort cloud population that is just 1--2 times larger than the outer Oort cloud \citep{dones04}. 

\begin{figure}[htb]
    \centering
    \includegraphics[scale=.75]{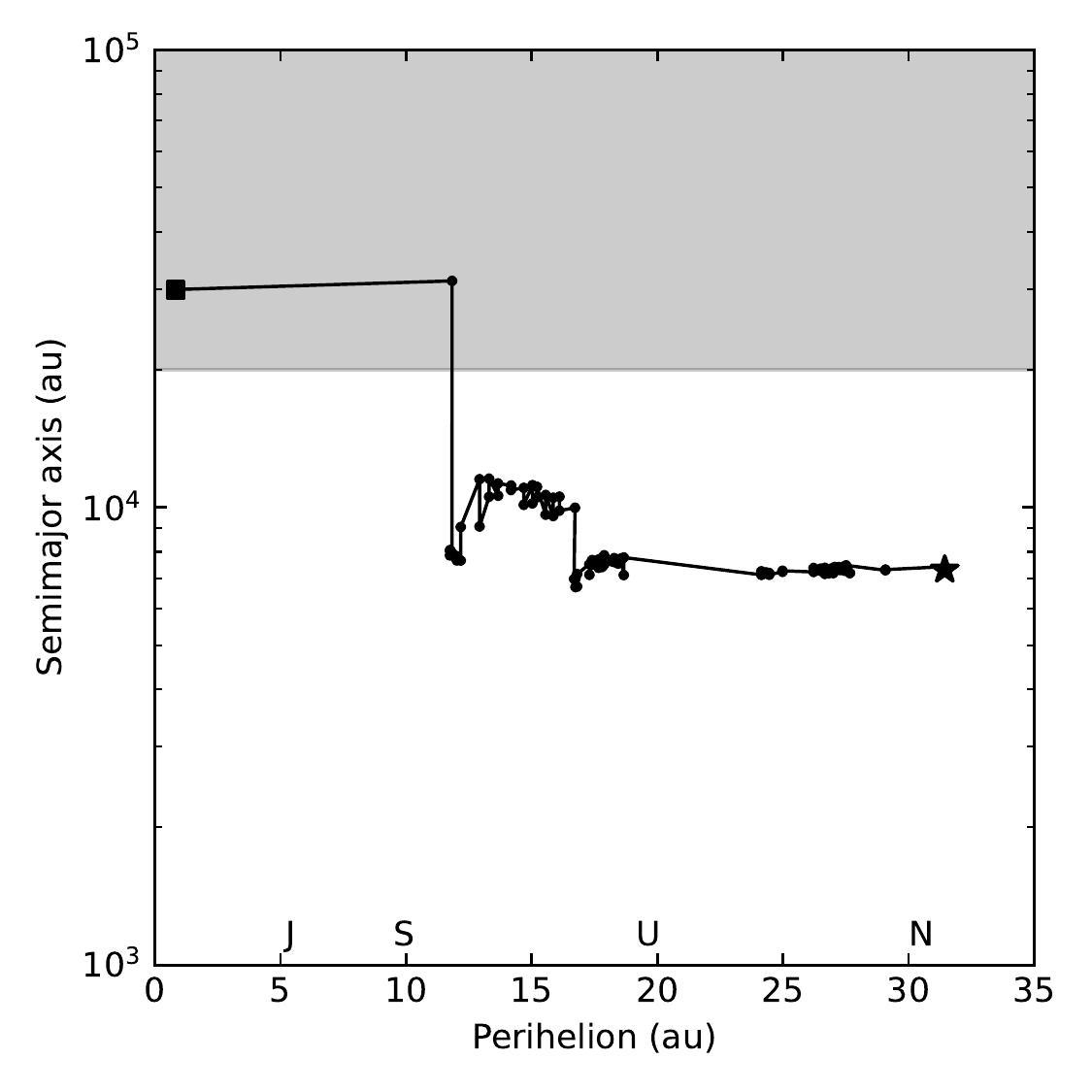}
    \caption{Plot of the orbital evolution of an Oort cloud body as it enters the solar system's planetary region. Data points mark the semimajor axis vs the perihelion each time the body enters the inner 50 au of the solar system. The evolution begins at the star data point and ends at the square data point. Letters mark the giant planet orbits along the $x$-axis. The shaded region marks Oort cloud semimajor axes historically assumed to supply observed LPCs, while unshaded semimajor axes had previously been assumed to be incapable of producing known LPCs.}
    \label{fig:KQ09Fig1}
\end{figure}

Since it was presumed that Jupiter and Saturn prevented the inner Oort cloud from contributing to the known catalog of LPCs \citep[the vast majority of which have perihelia of a few au or less;][]{mars08}, estimates of the inner Oort cloud population appeared to be completely model dependent. However, \citet{kaibquinn09} showed that objects from the inner Oort cloud can also evolve to become observable long-period comets. Illustrated in Figure \ref{fig:KQ09Fig1}, as the perihelia of inner Oort cloud bodies slowly march Sunward through the outer solar system, some can incur a set of planetary energy kicks that are too weak to eject them but sufficiently strong to inflate their semimajor axes above $\sim$20,000 au. At this point, their perihelia can evolve rapidly toward Earth in a single orbital period, allowing them to become discovered with semimajor axis that obscure their inner Oort cloud source region. 

This dynamical pathway was found to be roughly half as efficient at producing observable LPCs as the classic outer Oort cloud production \citep{kaibquinn09}. Thus, the current known population of LPCs implies that the inner Oort cloud population cannot be much greater than $\sim$two times the number of comets that had previously been estimated to occupy the outer Oort cloud. Otherwise the comet flux would be higher than what is observed. This rather modest inner Oort cloud population calls into question the significance of comet showers' relationship to mass extinctions. 

\section{\textbf{FUTURE DEVELOPMENTS}}
\label{sec:future}

\subsection{Rubin Observatory}\label{ss:rubin}

With a projected first light in the early to mid-2020s, the Vera C. Rubin Observatory promises to revolutionize our understanding of the small body populations of the solar system as it conducts the Legacy Survey of Space and Time (LSST) over ten years \citep{ivezic2019}. This observatory and survey is projected to discover tens of thousands of TNOs as well as provide the most extensive and uniform set of comet observations and discoveries in history. This will undoubtedly enhance our understanding of comets and their reservoirs. 

With an expected limiting magnitude of $\sim24.5$ covering all of the southern sky and the northern sky near the ecliptic plane, LSST's observations of the TNO populations will dwarf the present TNO catalog. To date, surveys reaching those limiting magnitudes have only covered small portions of the sky (e.g., the $\sim120$ deg$^2$ Outer Solar System Origins Survey; \citealt{Bannister:2018}), with all-sky surveys more typically limited to magnitudes brighter than $\sim22$ (e.g., PanStarrs; \citealt{Chambers:2016}).
LSST's combined wide sky coverage and deep limiting magnitude will dramatically increase the number of observed TNOs in the scattering and detached populations. 
This is critical to improving our models of both the intrinsic orbital and size distributions of these populations.
Current constraints on these populations have been limited by the fact that that these TNOs spend essentially all of their time at distances where they are un-observable. 
This limits the ability to directly de-bias observations (by weighting the observed objects by observability) to build a picture of the intrinsic population because the small current sample sizes make it likely we are missing examples from key regions of their orbital distributions. 
Likewise, forward modeling via the use of survey simulators to test models of the intrinsic scattering/detached TNO population against observations (see, e.g., \citealt{Lawler:2018ss}) must be done carefully because portions of the model phase space will be invisible, and thus un-testable.
The much larger observational sample expected from LSST, and the fact that it will come from a coherent survey with (hopefully) understood biases will lead to more detailed models of the scattering and detached populations.
The very wide latitude range of the observations (especially in the southern sky) will also provide strong new constraints on the inclination distribution of the TNOs.
This will provide better tests of solar system formation and giant planet migration scenarios, which will also inform ideas of where different kinds of comets formed. 
The improved size distribution measurements along with improved orbital distributions will also yield higher-fidelity models of the rate at which comets are supplied to the inner solar system and perhaps help inform models of cometary physical evolution. 

Rubin comet observations will dramatically improve our observational constraints on the Oort cloud. At present, our understanding of how cometary absolute magnitudes ($H_T$) relate to nuclear magnitudes is extremely uncertain. For each comet discovery, LSST will possess perhaps 50 separate temporally spaced observations made with the same instrument. This will allow us to precisely measure the photometric index of each comet as well as observe the comet at epochs when its brightness is less coma-dominated. In this manner, we can hope to measure nuclear magnitudes much more precisely and therefore estimate the size distribution of comets much more precisely. This in turn will enable much better mass estimates of the Oort cloud. 

In addition, the Rubin observatory will build up a catalog of LPCs with larger perihelia than most of those that have been previously discovered. This offers a new dynamical window to the Oort cloud because in order for a LPC to attain a perihelion within the terrestrial planet region ($q\lesssim5$ au), it must either already have possessed a large ($a\gtrsim20,000$ au) semimajor axis, or it must have attained one through previous encounter(s) with the giant planets at larger perihelia (see Figure \ref{fig:KRQ11}). Thus, the true semimajor axis distribution of Oort cloud bodies is obscured within our current sample of comets. As the sample of observed cometary perihelia approach and even exceed 10 au, the semimajor axis distribution of LPCs should change, and these changes will depend on the orbital distribution of bodies within the Oort cloud \citep{siltre18}. 

Rubin's anticipated TNO discoveries will also be immensely helpful in furthering our understanding of the Oort cloud. As discussed previously, highly inclined Centaurs are a class of objects that potentially originate from the Oort cloud. Their larger sizes and more distant perihelia offer a different set of constraints on the Oort cloud. Currently, $\sim$20 such bodies are known with perihelia beyond Saturn's orbit and inclinations over 45$^\circ$, and these have been detected with an amalgam of different surveys, making it difficult to account for observational biases within the current catalog. We should expect hundreds of such objects detected by the single LSST survey, giving us a less biased catalog of highly inclined centaurs, making it more straightforward to infer their ultimate dynamical source. 

LSST will also be capable of detecting Sedna-like bodies out to distances of over 100 au \citep{ivezic2019}. Thus, our sample of TNOs with large semimajor axes and perihelia very decoupled will increase by many fold. This will help us discern whether the Sun's cluster environment, a distant planet, or some other mechanism placed these objects on their current orbits. 

\subsection{Occultation Searches}\label{ss:occultation}

Observing $\sim$km-sized objects in-situ within the scattered disk or Oort cloud is not viable due to their extreme faintness. Moreover, the uncertainty of converting cometary brightness into nucleus size leaves our estimates of the population of $\sim$km-sized, or comet-sized, bodies within these reservoirs poorly constrained. While the crater size distribution observed by the New Horizons mission may offer some clues for the scattered disk (\citealt{Singer:2019, morb21}; see also Section~\ref{ss:obsSD}), stellar occultations may be the only tool to detect sizeable numbers of km-sized bodies as they orbit within the scattered disk and Oort cloud \citep{bail76, nihei07,Ortiz:2020}. When such objects pass in front of a star, the star's brightness will dim for a fraction of a second (typically, for a km-sized body at least tens of au from Earth). 

At present only a handful of serendipitous stellar outer solar system occultations have been observed. 
Using high-cadence stellar observations taken with HST's Fine Guidance Sensors, \citet{schlicht09} announced the discovery of a stellar occultation of a $r\simeq500$ m object at $\sim$45 au. With this single detection, they concluded that there was a deficiency of sub-km bodies within the Kuiper belt (and potentially the scattered disk) compared to a simple extrapolation of the power-law size distribution observed for TNO sizes above 100-km diameters. Analyzing additional HST guidance sensor data, \citet{schlicht12} discovered a second occultation by a similarly sized TNO at 40 au, which confirmed their previous conclusions about a deficiency of sub-km TNOs relative to an extrapolation of the $\sim$100-km TNO size distribution. The Taiwanese-American Occultation Survey (TAOS) also searched for such events by monitoring $\sim$500 stars at 5 Hz cadence for 7 years \citep{zhang13}. Although no occultation events were discovered, this non-detection placed an upper limit on the slope of the TNO size distribution below 90 km. A second generation survey is set to begin operations in 2022 (TAOS-II) that will monitor over 10$^4$ stars at 20 Hz cadence \citep{lehner12}. This occultation survey has a potential rate detection that is 100 times greater than its predecessor and will of course provide much tighter constraints on the TNO size distribution between $0.5\lesssim r \lesssim 30$ km. Although these constraints will largely be supplied by the classical Kuiper belt, they can likely be extrapolated to the scattered disk. 

Owing to its much larger average distance, occultation events by km-sized Oort cloud objects will generate much smaller decreases in stellar brightness \citep{nihei07}. This may place the detection of km-sized Oort cloud occultations beyond the realm of ground-based surveys \citep{nihei07}. 

\section{Summary}

New short-period comet and long-period comets are steadily produced from the large, distant reservoirs of the scattered disk and the Oort cloud, respectively. 
The scattered disk was generated early in the solar system's history via the dispersal of a massive primordial belt of planetesimals exterior to the formation region of the giant planets. 
This dispersal occurred when Neptune migrated outward by several au and the giant planets likely passed through an orbital instability \citep[e.g.][]{Malhotra:1993, Nesvorny:2018}. 
Less than one percent of these bodies were captured into the modern TNO populations, and an even smaller fraction now reside in the scattered disk. 
While some of these distant TNOs are actively scattering off of the giant planets today (the scattering TNOs, a fraction of which are destined to become JFCs), many are stored in mean motion resonances with Neptune and/or on orbits whose perihelia are dynamically detached from the planets. 
These detached orbits become unstable on timescales comparable to the solar system's age and seed the production of new scattering objects and JFCs. 
Our best current observational constraints imply a population of $\sim10^5$ scattering TNOs larger than $\sim100$~km in diameter \citep{Lawler:2018}. Extrapolations down to comet-sized objects are still very uncertain; models of the delivery of short period comets imply a few times $10^8$ scattering TNOs larger than $\sim2$~km in diameter \citep{nes17}.

Meanwhile, up to $\sim$5\% percent of the dispersed primordial belt bodies are captured into the Oort cloud at semimajor axes of $\sim$10$^{3-5}$ au \citep{dones04}. Most of this capture occurs during the first Gyr of the solar system's history. As the giant planets scatter these primordial belt bodies to larger semimajor axes, perturbations from passing field stars and the Galactic tide torque the perihelia of some of these bodies out of the planetary region, ending the scattering process and (at least temporarily) saving them from ejection. Although most dispersed bodies ultimately scatter off of Jupiter, this planet is more prone to ejecting bodies than emplacing them into the Oort cloud because the total orbital energy window of the Oort cloud is smaller than the typical energy kick delivered by Jupiter during a single encounter \citep{fernip84, dqt87}. Neptune and Uranus are much more efficient populators of the Oort cloud but are less likely to retain dynamical control of bodies throughout the scattering process. 

Simulations of the formation of the Oort cloud and scattered disk suggest that $\sim$10 times as many bodies should reside in the Oort cloud than in the scattering population \citep{dones04}. 
However, LPC observations suggest that the Oort cloud contains 7--8 $\times 10^{11}$ comet-size bodies (yielding an uncertain mass of 0.5--5 M$_{\oplus}$), which is a factor of $\sim$50--1000 times larger than scattered disk population estimates \citep{brasmorb13, lev10}.  This potentially represents a major discrepancy between formation models and comet observations, but this population ratio estimate relies on a highly uncertain conversion of cometary magnitude to nucleus size. 

Simulations have shown that the structure of the Oort cloud is highly sensitive to the Sun's dynamical history within the Galaxy, in particular its stellar birth cluster and the degree of radial migration it has undergone within the Milky Way \citep{fern97, kaib11}. Both of these effects can push the Oort cloud's range of semimajor axes closer to the Sun, potentially as close as several hundred au. The exact inner edge of the Oort cloud is also very dependent the handful of closest stellar encounters that have occurred over the Sun's history, and the process of setting the inner edge is therefore quite stochastic \citep{kaib11}. While it has been previously speculated that the number of comet-sized bodies in the interior 20,000 au of the Oort cloud could be many times the number that are exterior to 20,000 au, we now know that LPCs can be generated from Oort cloud semimajor axes as small as $\sim$5000 au \citep{kaibquinn09}. This finding places an upper limit of $\sim$10$^{12}$ comet-sized bodies orbiting beyond $\sim$5000 au in the Oort cloud. 

To date, much of our understanding of the population and orbital structure of the Oort cloud relies upon observations of comets passing near Earth. 
In the coming decade, the Vera C. Rubin Observatory will increase the inventory of small solar system bodies by many fold \citep{ivezic2019}. 
Its discoveries will include a sample of comets discovered and observed at large ($\gtrsim$10 au) perihelion and heliocentric distance that will have experienced less thermal processing (by the Sun) and dynamical processing (by Jupiter) compared to the historical comet catalog. 
In addition, we expect LSST to increase the inventory of observed TNOs by an order of magnitude, including discovering many more objects at high ($\gtrsim$ 45$^{\circ}$) inclinations and with large ($\sim$40--100 au) perihelia. 
Finally, we anticipate the scope and sensitivity of stellar occultation searches to advance in the coming decade, and this should provide new, comet-independent constraints on the number of km-sized bodies in the scattered disk \citep{lehner12}. However, a similar effort for the Oort cloud may require a space-based campaign \citep{nihei07}. These anticipated discoveries will provide new windows into the population and structure of the scattered disk and the Oort cloud, whose properties offer clues to the giant planet's orbital evolution as well as the Sun's dynamical history. 



\vskip .5in
\noindent \textbf{Acknowledgments.} \\

We thank reviewer Julio Fern\'{a}ndez and another anonymous reviewer for comments and suggestions that greatly improved this review. NAK thanks the Department of Astronomy at Case Western Reserve University for hosting him as a visitor during this work's preparation. NAK acknowledges support from NSF CAREER award 1846388 and NASA Emerging Worlds grant 80NSSC18K0600. KV acknowledges support from NSF (grant AST-1824869) and NASA (grants 80NSSC19K0785 and 80NSSC21K0376).

\bibliographystyle{sss-three.bst}
\bibliography{refs.bib,refs-from-ads.bib}

\end{document}